\documentclass[11pt, oneside]{article}   
\usepackage{geometry}                		
\usepackage{amsmath}						
\usepackage{amssymb}
\usepackage{subcaption}
\pdfoutput=1
\usepackage{jheppub}

\usepackage{epsfig} 
\usepackage{psfrag}
\usepackage{amsmath}
\usepackage{amssymb}
\usepackage{color}
\usepackage{graphicx}
\usepackage{hyperref} 
\usepackage{enumitem}
\usepackage{braket}
\usepackage{relsize}
\usepackage{float}
\usepackage{color}

\usepackage{placeins}

\usepackage{comment}

\begin{document}
 
\title{Operator complexity: a journey to the edge of Krylov space} 

\author[a]{E. Rabinovici,} 
\emailAdd{eliezer@mail.huji.ac.il}

\author[b]{A. S\'{a}nchez-Garrido,}
\emailAdd{Adrian.SanchezGarrido@unige.ch}

\author[a]{R. Shir}
\emailAdd{ruth.shir@mail.huji.ac.il}

\author[b]{and J. Sonner}
\emailAdd{Julian.Sonner@unige.ch}

\affiliation[a]{Racah Institute of Physics,
The Hebrew University, Jerusalem 9190401, Israel}
\affiliation[b]{Department of Theoretical Physics, University of Geneva, 24 quai Ernest-Ansermet, 1214 Gen\`eve 4, Switzerland}

\abstract{
  Heisenberg time evolution under a chaotic many-body Hamiltonian $H$ transforms an initially simple operator into an increasingly complex one, as it spreads over Hilbert space. Krylov complexity, or `K-complexity', quantifies this growth with respect to a special basis, generated by $H$ by successive nested commutators with the operator. In this work we study the evolution of K-complexity in finite-entropy systems for time scales greater than the scrambling time $t_s>\log (S)$. We prove rigorous bounds on K-complexity as well as the associated Lanczos sequence and, using refined parallelized algorithms, we undertake a detailed numerical study of these quantities in the SYK$_4$ model, which is maximally chaotic, and compare the results with the SYK$_2$ model, which is integrable. While the former saturates the bound, the latter stays exponentially below it. 
  We discuss to what extent this is a generic feature distinguishing between chaotic vs. integrable systems.
}

\date{September 2020}


\maketitle

\section{Introduction}
The concepts of state and operator complexity lie at the intersection of quantum information, condensed matter physics, quantum field theory (QFT) and black hole physics. One is thus faced with a diversity of concepts and methods on how to precisely define and actually measure complexity. The identification of very large time scales in black hole physics and corresponding phenomena in QFT (via the AdS/CFT correspondence) \cite{Maldacena:2001kr, Barbon:2003aq, Dyson:2002pf,  Susskind:2014rva, Stanford:2014jda, Barbon:2014rma, Cotler:2016fpe, Brown:2017jil} has motivated the search for a form of complexity whose time dependence has the following distinctive characteristics: In fast-scrambling systems with finite entropy $S$, complexity should grow exponentially in time until $t_s \sim \log (S)$ (known as the scrambling time) when it reaches a value of order ${\cal O}(S)$; it then switches to a linear-in-time behavior until times of order $\exp({\cal O}(S))$ when it settles around a plateau of order $\exp({\cal O}(S))$; finally, after times of order $\exp(\exp({\cal O} (S)))$ it is expected to start performing the Poincar\'e dance. The growth of the Einstein-Rosen bridge was connected in \cite{Susskind:2014rva} with complexity growth of the quantum state in the dual CFT. A notion of complexity known as Krylov complexity, or K-complexity, was introduced in \cite{Parker:2018yvk} as a useful diagnostic of quantum chaos in the thermodynamic limit. In \cite{Barbon:2019wsy} it was argued for the first time that K-complexity of generic operators in finite-size systems exhibits the profile of complexity described above. This paper reports on a complete numerical analysis of K-complexity for all time scales, including scrambling and well beyond, in concrete models demonstrating the above-described time-dependence. This is the first time that such large time scales are studied with numerical methods, and in concrete models. We uncover non-perturbative effects which were not initially anticipated. The study overcame known numerical instabilities and employed powerful algorithms executed on large computing clusters.

Roughly speaking the time evolution of complexity quantifies how quickly a reference operator grows under Heisenberg evolution in a pre-defined basis. K-complexity, denoted $C_K$, measures this growth with respect to the Krylov basis which, as we shall see below, is well-adapted to capture Heisenberg evolution efficiently. K-complexity depends on the Hamiltonian of the system and a chosen reference operator. It has an advantage over other notions of complexity: its definition doesn’t require an arbitrary tolerance parameter. For circuit complexity \footnote{Circuit complexity of the SYK$_4$ model, the main work-horse of this paper, has previously been studied in \cite{Garcia-Alvarez:2016wem,babbush2019quantum}.}, to cite but one example, such a tolerance parameter must be introduced, and its presence is crucial in establishing its boundedness \cite{Kitaev_1997, nielsen_chuang_2010, dawson2005solovaykitaev}, whereas K-complexity is naturally bounded from above \footnote{Operator size complexity \cite{Roberts:2018mnp} shares this advantage of not requiring a tolerance parameter, but differs from $C_K$ in other crucial respects. For example, it is directly bounded by the total system size, while the upper bound for K-complexity scales exponentially with system size.} \cite{Barbon:2019wsy}. It is known to be bounded by the dimension of the Hilbert space of operators; below we prove a stronger bound.

$C_K$ provides a fine-grained notion of complexity, as it manifestly distinguishes all linearly independent operators up to a given, fixed length. These features make K-complexity a natural choice for quantifying the evolution of operators at late times (see e.g. Sec. 2.1 in \cite{Altland:2020ccq} for a careful definition of chaotic time scales) which is this paper’s focus.

In studying K-complexity at these time scales one should either explicitly construct, or obtain bounds on, the {\it full} sequence of Lanczos coefficients, $b_n$, which characterize the Krylov basis of an operator. These coefficients arise in the process of orthonormalizing\footnote{The process of orthonormalization requires a choice of inner product. In this work we use the standard Frobenius inner-product (see Section \ref{Sec: Lanczos algorithm}). For other choices of inner product inherited from finite-temperature correlation functions see discussions in \cite{Recursion_Method, Parker:2018yvk}.} the Krylov basis \cite{Lanczos1950AnIM, Recursion_Method}. In \cite{Parker:2018yvk} it was hypothesized that in a chaotic quantum system in the thermodynamic limit, the Lanczos coefficients grow at a linear rate. In \cite{Barbon:2019wsy} it was suggested that, in finite chaotic systems, the Lanczos-sequence could exhibit plateau-like features over a certain range. The study of finite chaotic systems reported in this paper constructs the full Lanczos sequence to reveal a non-perturbative decay of this plateau, making an initially small but eventually persistent correction to the “cliff-ended plateau" conjectured in \cite{Barbon:2019wsy}. We refer to the resulting profile as “the Descent”.

We study K-complexity in the maximally chaotic SYK$_4$ model \cite{Sachdev:1992fk, Kitaev:2015, Sachdev:2015efa, Maldacena:2016hyu}, which is by now well-established as a toy model displaying key chaotic properties of black holes, including late-time spectral chaos \cite{Garcia-Garcia:2016mno, Cotler:2016fpe}. We also study the quadratic SYK$_2$ model, which is integrable, showing Poisson statistics \cite{Garcia-Garcia:2017bkg, Garcia-Garcia:2020cdo, Haque_2019}.

The structure of this paper is as follows: In Section \ref{Sec: Krylov space} we describe the Krylov space and an analytical upper bound on its dimension. In Section \ref{Sec: Integrable systems} we discuss the Krylov space dimension for integrable systems, focusing on SYK$_2$. In Section \ref{Sec: Lanczos algorithm} the Lanczos algorithm by which an orthonormal basis for Krylov space as well as the Lanczos-sequence are constructed, is described. In Section \ref{Sec: KC and KS} we present the definitions of K-complexity and K-entropy as well as expectations for their saturation values.  Section \ref{Sec: Numerical results} shows our numerical results for the Lanczos-sequences, K-complexity and K-entropy for complex SYK$_4$ systems with $8,9$ and $10$ fermions. In Section \ref{Sec: Discussion} we summarize the main points of our results, remark on some of the differences between K-complexity and circuit complexity and on possible holographic aspects. Appendix \ref{Appx-RMT} presents numerical results for the Lanczos-sequence in Random Matrix Theory. Appendix \ref{Appx-SYK2} discusses the Krylov basis for SYK$_2$ and Appendix \ref{Appx-algorithms} shows explicitly the numerical algorithms used to produce the data in the paper.

\section{Time evolution of operators and physical properties of Krylov space} \label{Sec: Krylov space}

We begin by providing the definition of Krylov space and stressing its relation to the structure of the spectrum of the system under consideration (and hence to its integrable or chaotic character). Consider a Hilbert space $\mathcal{H}$, of dimension $\dim\left(\mathcal{H}\right) = D$, with associated Hilbert space of linear operators ${L}\left(\mathcal{H}\right)\equiv \widehat{\mathcal{H}}$, acting on $\mathcal{H}$ and satisfying $\dim(\widehat{\mathcal{H}})=D^2$. Now consider a system whose dynamics are described by a certain (hermitian) Hamiltonian $H\in\widehat{\mathcal{H}}$, and an observable $\mathcal{O}\in \widehat{\mathcal{H}}$.

The associated Krylov space \cite{Kry31} is defined as the minimal subspace of operator space that contains the time evolution of $\mathcal{O}$ at all times. 

The time evolution of $\mathcal{O}$ is given by
\begin{equation}
    \centering
    \label{BCH-Liouvillian-def}
    \mathcal{O}(t) = e^{iHt}\mathcal{O} e^{-iHt} = e^{i\mathcal{L} t}\mathcal{O}
\end{equation}
where the Liouvillian operator is defined as $\mathcal{L}\equiv \left[H,\cdot\right]$. Thus the Krylov space is the linear span of all nested commutators of the Hamiltonian with the operator:
\begin{equation}
    \centering
    \label{Krylov-definition_maintext}
    \mathcal{H}_\mathcal{O} = \text{span}\left\{\mathcal{L}^n\mathcal{O}\right\}_{n=0}^{+\infty}= \text{span}\left\{\mathcal{O},\, [H,\mathcal{O}],\, [H,[H,\mathcal{O}]], \dots\right\}.
\end{equation}
Equivalently, $\mathcal{H}_\mathcal{O}$ is the minimal invariant subspace of the Liouvillian that contains $\mathcal{O}$. We denote $\dim\left(\mathcal{H}_\mathcal{O}\right)\equiv K$. 

$K$ is determined by studying the cardinality of the maximal set of linearly independent objects of the form $\mathcal{L}^n\mathcal{O}$, which can be found by
computing the rank of
\begin{equation}
    \label{BCH-basis-rank}
    \left(\begin{matrix}
         \mathcal{O}, &
        \mathcal{LO}, &
         \mathcal{L}^2\mathcal{O},&
        \dots &
        \mathcal{L}^n\mathcal{O}, &
        \dots
    \end{matrix}\right)^T\,,
\end{equation}
in a convenient matrix representation, which we now define.
Let us choose the basis $|\omega_{ab})\equiv\ket{E_a}\bra{E_b}$, on $\widehat{\cal H}$, which is naturally induced by the eigenbasis of the Hamiltonian on ${\cal H}$. Then each nested commutator takes the form:
\begin{equation}
    \centering
    \label{Op-phases}
    \mathcal{L}^n\left|\mathcal{O}\right) =\delta_{n0} \sum_{a=1}^D O_{aa}|\omega_{aa})+\sum_{\substack{a,b = 1 \\ a\neq b}}^D  O_{ab}\,  \omega_{ab}^n \, |\omega_{ab}) 
\end{equation}
where we have defined the phases \footnote{We shall refer to energy differences as \textit{phases}, since they appear as such in the moment expansion of the two-point function of $\mathcal{O}$.} $\omega_{ab}:= E_a-E_b$, which are eigenvalues of the Liouvillian acting on $|\omega_{ab})$. Note that the $\omega$'s related to pairs of the form $(a,a)$ are zero, $\omega_{aa} = 0$ for all $a=1,...,D$. Displaying the coordinates of (\ref{Op-phases}) as rows one can construct a matrix representation of (\ref{BCH-basis-rank}), which turns out to be a Vandermonde matrix. The rank of this matrix is at most $D^2$, as it has $D^2$ columns, so we shall keep the same number of rows to make it a square Vandermonde matrix:

\begin{equation}
    \centering
    \label{Vandermonde_matrix_explicit}
    \begin{pmatrix}
    O_{11} & O_{22} & \dots & O_{DD} & O_{12} & O_{13} & \dots & O_{D-1,D} \\
    
    0 & 0 & \dots & 0 & O_{12}\; \omega_{12} & O_{13}\; \omega_{13} & \dots & O_{D-1,D}\;\omega_{D-1,D} \\
    
    0 & 0 & \dots & 0 & O_{12}\; \omega_{12}^2 & O_{13}\; \omega_{13}^2 & \dots & O_{D-1,D}\;\omega_{D-1,D}^2 \\
    \vdots & \vdots & \ddots & \vdots & \vdots & \vdots & \ddots & \vdots \\
    0 & 0 & \dots & 0 & O_{12}\; \omega_{12}^{D^2-1} & O_{13}\; \omega_{13}^{D^2-1} & \dots & O_{D-1,D}\;\omega_{D-1,D}^{D^2-1}
    \end{pmatrix} ~.
\end{equation}
To find its actual rank, we can compute its determinant, which is given by:
\begin{equation}
    \centering
    \label{Det-v2}
    \Delta\left(\left\{\omega_{ab}\right\}\right)\prod_{i,j=1}^{D} O_{ij}\,,
\end{equation}
where $\Delta\left(\left\{\omega_{ab}\right\}\right)$ is the Vandermonde determinant of the phases in the matrix. The expression (\ref{Det-v2}) will be zero if any of the phases are degenerate, and also if any of the matrix elements in the energy basis vanish, so the corresponding columns should be removed in order to be left with a matrix of maximal rank. Hence, the Krylov dimension $K$ can be estimated using the following algorithmic prescription: $K$ is equal to the number of distinct phases corresponding to the indices of non-zero matrix elements of the operator in the energy basis. The zero phase $\omega_{aa}=0$ is always at least $D$ times degenerate and therefore, since it can only be counted once, the Krylov dimension is bounded by:
\begin{equation}
    \label{Krylov-bound}
    1\leq K \leq D^2-D+1
\end{equation}
for any non-vanishing operator.

In general, if the operator $\mathcal{O}$ has a non-vanishing projection over several eigenstates of the Liouvillian that share the same degenerate eigenvalue, then this phase only contributes one dimension to the Krylov space. So a legitimate question is to wonder what particular linear combination of those eigenstates is actually contained in the Krylov space. In order to answer this, one can consider the form of the operator $\mathcal{L}^n\left|\mathcal{O}\right)$ given in (\ref{Op-phases}). Suppose that, for some set $I$ of pairs of indices, the eigenvalue is degenerate:
\begin{equation}
    \centering
    \label{Deg-phase}
    \mathcal{L}\left|\omega_{ab}\right) = \omega\left|\omega_{ab}\right)\;\;\;\forall (a,b)\in I
    \end{equation}
i.e. $\omega_{ab}=\omega$ for all $(a,b)\in I$. Assume also that $\omega_{ab}\neq\omega$ for any $(a,b)\notin I$. Inserting (\ref{Deg-phase}) in (\ref{Op-phases}) one finds:
\begin{equation}
    \centering
    \mathcal{L}^n\mathcal{O}=\omega^n \sum_{\left(a,b\right)\in I}\,O_{ab}\left|\omega_{ab}\right)+\sum_{\left(a,b\right)\notin I}O_{ab}\,\omega_{ab}^n\left|\omega_{ab}\right).
\end{equation}
It is manifest now that the direction of the $\omega$-eigenspace that contributes to the Krylov space is precisely the projection of the operator $\mathcal{O}$ over such an eigenspace:

\begin{equation}
    \centering
    \label{eigenspace-contrib}
    \left|\mathcal{K}_\omega\right):= \sum_{\left(a,b\right)\in I}\,O_{ab}\left|\omega_{ab}\right)~.
\end{equation}
Let`s call $\left|\mathcal{K}_\omega\right)$ the \textit{eigenspace representative} for the phase $\omega$. The structure of the Krylov space is now fully understood: Each eigenspace of the Liouvillian over which the operator $\mathcal{O}$ has a non-vanishing projection contributes one Krylov dimension, through a linear combination of the basis of the eigenspace of the form (\ref{eigenspace-contrib}). We can thus redefine the Krylov space as:
\begin{equation}
\centering
\label{Krylov-eigenspaces}
\mathcal{H}_\mathcal{O}=\text{span}\left\{\left|\mathcal{K}_\omega\right),\;\;\;\omega\in \sigma(\mathcal{L})\right\}
\end{equation}
where $\sigma(\mathcal{L})$ denotes the spectrum of $\mathcal{L}$. 
Finally, the Krylov dimension $K$ is simply equal to the number of non-zero eigenspace representatives $\left|\mathcal{K}_\omega\right)$.

For instance, if one considers a system whose Liouvillian has a spectrum with no degeneracies (other than the unavoidable null phases) and an operator that is dense in the energy basis, the Krylov dimension will be maximal, $K=D^2-D+1$. The only source of degeneracy in the phases will be the universal one due to diagonal phases $\omega_{aa}=0$. We can now note explicitly that the part of the operator algebra that is left out of the Krylov space belongs to the space subtended by the projectors on the energy eigenstates, since those correspond to zero phases, and the only combination of them that contributes to $\mathcal{H}_\mathcal{O}$ is the representative of the $\omega=0$ eigenspace, in this case:
\begin{equation}
    \centering
    \label{eigenspace-0-Krylov}
    \left|\mathcal{K}_0\right) = \sum_{a=1}^D O_{aa}\left|\omega_{aa}\right)=\sum_{a=1}^D O_{aa}\ket{E_a}\bra{E_a}\equiv\sum_{a=1}^DO_{aa}\mathcal{P}_a.
\end{equation}

To summarize, $K$ is determined by exploiting the advantages of the basis in operator space induced by the eigenbasis of the Hamiltonian; crucially, the determinant of a suitable matrix containing the representation of these nested commutators in the given basis reduces to a Vandermonde determinant of energy differences. The following algorithmic prescription is derived: $K$ is equal to the number of eigenspaces of the Liouvillian over which $\mathcal{O}$ has non-zero projection. The eigenvalues of the Liouvillian are precisely all possible energy differences, $\omega_{ab}=E_a-E_b$. The zero phase is always at least $D$ times degenerate, thus the Krylov dimension is bounded by (\ref{Krylov-bound}),
for any non-zero operator. The more degeneracies there are in the spectrum of the Liouvillian, the lower will be the Krylov dimension. The expectation is that the spectrum of a chaotic system will not feature degeneracies apart from those induced by the presence of extra symmetries; we therefore conjecture that typical \footnote{Along the lines of ETH \cite{PhysRevA.30.504, PhysRevA.43.2046, Srednicki_1999, DAlessio:2016rwt}, we expect typical observables in chaotic systems to be dense in the energy basis (as is the case of localized operators in local chaotic systems). Regardless of the structure of the spectrum, one can always choose a fine-tuned operator having a sparse matrix representation in the energy basis, e.g. an eigenspace projector $\ket{E}\bra{E}$; such a candidate will indeed have a very small associated Krylov space, but won't generally fulfil the requirements to be considered a \textit{typical} operator. Conserved charges are also special operators, since they commute with the Hamiltonian, and therefore their Krylov space is always one-dimensional.} operators in chaotic systems saturate the upper bound in (\ref{Krylov-bound}); for integrable systems we expect $K$ to be significantly smaller than the bound. We have confirmed both expectations by studying the chaotic SYK$_4$ model (see numerics below) as well as RMT (see Appendix \ref{Appx-RMT}), and the integrable SYK$_2$ model, see Section \ref{Sec: Integrable systems} below. 

\section{Integrable systems and SYK$_2$} \label{Sec: Integrable systems}
We have conjectured that the Krylov space for the time evolution of a typical operator for integrable systems will be significantly smaller than the upper bound in (\ref{Krylov-bound}).
Before verifying our conjecture in the case of SYK$_2$, we verify it in the simpler example of the quantum harmonic oscillator $H= \omega(a^\dagger a +1/2)$ with the position operator $\hat{x}=\sqrt{\frac{1}{2\omega}}(a+a^\dagger)$, for which the Krylov space dimension is $K=2$, even though the Hilbert space is infinite\footnote{Accordingly, the Lanczos-sequence contains a single element $b_1=\omega$ (see next section for the definition of the Lanczos-sequence).}. In this case, the smallness of the Krylov space is due to the fact that the position operator has a non-vanishing projection over only two eigenspaces of the Liouvillian (those corresponding to the energy differences $\omega$ and $-\omega$).

We now test our conjecture in SYK$_2$, an integrable system with more structure than the harmonic oscillator.
For simplicity, let us use the Majorana (real) version of the model \cite{Kitaev:2015,Maldacena:2016hyu}:
\begin{equation}
    \centering
    \label{MajoranaSYK-q2}
    H = i \sum_{1\leq i < j \leq L}m_{ij}\,\chi_i \chi_j
\end{equation}
where the coupling strength $m_{ij}\in \mathbb{R}$ is antisymmetric, $m_{ij}=-m_{ji}$, and each independent matrix element is drawn from a Gaussian distribution with zero mean and variance $\mathbb{E} (m_{ij}^2) = m^2/L$.
The number of sites is even, $L=2M$, and the Majorana fermions satisfy the relations
    $\left\{\chi_i,\chi_j\right\}=\delta_{ij},\;\;\;\chi_i=\chi_i^\dagger$.

The key point in the SYK$_2$ case is the fact that operators do not grow after commutation with the Hamiltonian, as also observed by \cite{Roberts:2018mnp, Carrega:2020mah}. Hence the subspace containing their time evolution will be, at most, that subtended by operators of fixed equal size. For example if the operator of study is a Majorana on some site with index $A$, $\mathcal{O}\equiv\chi_A$, commutation with the Hamiltonian will give $\left[\chi_i\chi_j,\chi_A\right] = \delta_{Aj}\chi_i-\delta_{Ai}\chi_j$, which is again a single-site operator (more details on the construction of the Krylov space in this case are given in Appendix \ref{Appx-SYK2}). For one-site operators, this implies an upper bound on $K$:
    \begin{equation}
        \centering
        \label{q2-KDIM-main}
        K\leq L \sim \log D \ll D^2-D+1 ~.
    \end{equation}
It should be possible to reach this conclusion by studying the degeneracy structure of the spectrum of SYK$_2$, which features Poisson level spacing statistics, and it is natural to expect other integrable systems to have a small Krylov space, due to the expected degeneracies in the spectrum of the Liouvillian and to the sparseness of simple operators in the energy basis.
In general, interacting integrable systems feature Poisson level spacing statistics, implying, at least, the existence of quasi-degenerate levels in the spectrum of the Hamiltonian, and therefore also in that of the Liouvillian. We expect these systems to behave as if the degeneracies were exact to a good degree of approximation, hence admitting a description in terms of an effectively smaller Krylov space. This argument applies, and accounts for a significant effective reduction of Krylov space, even if the operator is not sparse in the energy basis, as long as the invariant subspace of the Liouvillian over which it has a non-trivial projection contains degenerate or quasi-degenerate levels. All this leads us to suggest that integrable systems have a lower effective Krylov dimension compared with chaotic ones. 
As a first step in trying to elevate this conjecture based on a few concrete examples as well as qualitative ones, we estimate the minimal length of time for which the difference in the dimension of the explored Krylov space is large.
Since quasi-degenerate energy levels are separated by less than the mean level spacing $\Delta \sim \Gamma e^{-S}$, where $\Gamma$ is a relevant energy scale of the system such as the total spectral width, we can estimate the lowest time scale until which the Krylov dimensions of chaotic vs. integrable (with quasi-degeneracies) systems remain significantly different, to be of order of the Heisenberg time, i.e of order $e^{O(S)}$. For infinite systems this lasts forever, while for large finite systems this is way above the thermalization or scrambling time scales. The question of what ultimately generically happens between the Heisenberg time scale and the Poincar\'e time scale requires further research.
To our knowledge, bounds like (\ref{Krylov-bound}) and (\ref{q2-KDIM-main}) haven't been noted previously in the literature.

\section{Lanczos algorithm} \label{Sec: Lanczos algorithm}

Once the Krylov space adapted to time evolution of an operator with a constant Hamiltonian is identified, one would like to construct an orthonormal basis for it, given a certain scalar product $\left(\cdot|\cdot\right)$ on operator space. This is achieved with the Lanczos algorithm, which is a particularization of the Gram-Schmidt procedure:
\begin{enumerate}
    \item set $b_0\equiv0$ and $\left|\mathcal{O}_{-1}\right)\equiv0$
    \item $\left|\mathcal{O}_0\right) =\frac{1}{\sqrt{\left(\mathcal{O}|\mathcal{O}\right)}} \left|\mathcal{O}\right)$
    \item for $n\geq 1$: $\left|\mathcal{A}_n\right) = \mathcal{L}\left|\mathcal{O}_{n-1}\right) - b_{n-1}\left|\mathcal{O}_{n-2}\right)$
    \item set $b_n = \sqrt{\left(\mathcal{A}_n|\mathcal{A}_n\right)}$
    \item if $b_n=0$ stop; otherwise set $\left|\mathcal{O}_n\right) = \frac{1}{b_n}\left|\mathcal{A}_n\right)$ and go to step 3.
\end{enumerate}

In this work we make use of the Frobenius product $\left(A|B\right)=\frac{1}{D}\text{Tr}\left[A^\dagger B\right]$. The algorithm will construct an orthonormal set $\left\{\mathcal{O}_n\right\}_{n=0}^{K-1}$, the \textit{Krylov basis}, and the \textit{Lanczos coefficients} $\left\{b_n\right\}_{n=1}^{K-1}$.

Each Lanczos step produces an element $\left|\mathcal{A}_n\right)$ orthogonal to all previous $\left|\mathcal{O}_m\right)$ with $m<n$, so it is either zero or a new direction in Krylov space. For $n<K$, $\left|\mathcal{A}_n\right)\neq 0$ because the set that is being orthogonalized with this procedure has rank $K$ (in particular, $\left|\mathcal{A}_n\right)$ contains terms with up to $n$ nested commutators of $H$ with $\mathcal{O}$). However, $\left|\mathcal{A}_K\right)$ is orthogonal to $\left\{\mathcal{O}_n\right\}_{n=0}^{K-1}$, which is already a complete orthonormal basis of $\mathcal{H}_\mathcal{O}$, so it must therefore vanish, just as $b_K=\sqrt{\left(\mathcal{A}_K|\mathcal{A}_K\right)}=0$. We conclude that the Lanczos algorithm must terminate by hitting a zero once all directions in Krylov space have been exhausted. This is accounted for in Step 5 above.

The representation of the Liouvillian over the Krylov space in such a basis simplifies to a tridiagonal matrix $\left(\mathcal{O}_m\right|\mathcal{L}\left|\mathcal{O}_n\right)=T_{mn}$, whose entries are given by the Lanczos coefficients:
\begin{equation}
\centering
\label{L-Tridiagonal}
     T=\begin{pmatrix}
        0 & b_1 & 0 & 0 & \dots & 0\\
        b_1 & 0 & b_2 & 0 &\dots & 0 \\
        0 & b_2 & 0 & b_3 & \dots & 0 \\
        \vdots &  & \ddots & & \ddots & \vdots \\
         & & & & \ddots & b_{K-1}\\
        0 & 0 & 0 & \dots & b_{K-1} & 0
    \end{pmatrix}.
\end{equation}
The eigenvalues of this matrix are all non-degenerate, and given precisely by the phases corresponding to the eigenspaces of the Liouvillian that span the Krylov space (see Section \ref{Sec: Krylov space}).

The original Lanczos algorithm described above is known to suffer from numerical instabilities which can be overcome using the re-orthogonalization algorithms FO and PRO \cite{PRO, Parlett_book} described in Appendix \ref{Appx-algorithms}.

\section{K-complexity and K-entropy} \label{Sec: KC and KS}
Using the sequence of Lanczos coefficients (which we will also call the $b$-sequence), one can reduce the analysis of the time-evolution of an operator $\mathcal{O}$ into the solution of a differential recurrence equation. The time-evolving operator can be expanded in the Krylov basis as:
\begin{equation}
    \label{Expansion-Krylov-basis}
    \left|\mathcal{O}(t)\right)=\sum_{n=0}^{K-1}i^n\varphi_n(t)\left|\mathcal{O}_n\right) ~,
\end{equation}
where $\varphi_n(t)$ are time-dependent coefficients which describe how the operator is distributed over the Krylov basis (they may be thought of as the ``wavefunctions" in $n$). Given the Heisenberg equation $\frac{d\mathcal{O}}{dt}=i[H,\mathcal{O}]$, they satisfy the differential recurrence equation
\begin{equation}
    \label{difrec}
    \dot{\varphi}_n(t)=b_n\varphi_{n-1}(t)-b_{n+1}\varphi_{n+1} ~.
\end{equation}
Here, $\varphi_n(0)=\delta_{n0}$, since for a normalized operator $\mathcal{O}(0)=\mathcal{O}_0$. We set $\varphi_{-1}(t)\equiv 0$ in order to make the recurrence (\ref{difrec}) consistent with the definition (\ref{Expansion-Krylov-basis}). Also, from unitarity $\sum_{n=0}^{K-1}|\varphi_n(t)|^2=1$. The Lanczos coefficients ${\left\{ b_n \right\}}_{n=1}^{K-1}$ can be understood as hopping amplitudes for the initial operator $\mathcal{O}_0$ to explore the ``Krylov chain" and the functions $\varphi_n(t)$ can be visualized as wave-packets travelling on it \cite{Parker:2018yvk}. 

We now display two quantities which highlight broad features of the distribution $\varphi_n(t)$, viz.
\begin{itemize}
    \item \textbf{K-complexity}, which computes the average position of the distribution on the ordered Krylov basis:
    \begin{equation}
        \centering
        \label{K-Complexity}
        C_K(t)=\sum_{n=0}^{K-1}n|\varphi_n(t)|^2.
    \end{equation}
    \item \textbf{K-entropy}, which computes how randomized the distribution is:
    \begin{equation}
        \centering
        \label{K-Entropy}
        S_K(t)=-\sum_{n=0}^{K-1} |\varphi_n(t)|^2\log |\varphi_n(t)|^2.
    \end{equation}
\end{itemize}
Assuming that at very late times the operator is evenly distributed over Krylov space, there exists a saturation time $t_{sat}$ for which 
\begin{equation} \label{phi_sat}
   |\varphi(t\geq t_{sat})|^2\sim \frac{1}{K} ~.
\end{equation}
We can then get a rough estimate for the values of K-complexity and K-entropy at very late times by plugging (\ref{phi_sat}) into the formulas for $C_K(t)$ and $S_K(t)$:
\begin{equation}
    \label{KCsat}
    C_{K}(t\geq t_{sat})\sim\frac{1}{K} \frac{K(K-1)}{2}\sim \frac{K}{2} ,
\end{equation}

\begin{equation}
    \label{KEsat}
    S_K(t\geq t_{sat})\sim-K \frac{1}{K} \log(1/K)= \log(K).
\end{equation}
Since for chaotic systems $K\sim D^2$ and in general $D\sim e^S$, where $S$ is the entropy of the system (in the sense of ``log of the number of states"), we find that the saturation value of K-Entropy is essentially of order $S$, while the saturation value of K-complexity is of order $e^{2S}$. If $C_K(t)$ grows linearly after scrambling, the saturation time will be roughly $t_{sat}\equiv t_K\sim e^{2S}$, in agreement with the expectation in \cite{Barbon:2019wsy}. These properties are confirmed in our numerical results.

\section{Numerical results} \label{Sec: Numerical results}
In the following we present numerical results for complex SYK$_4$, whose Hamiltonian is schematically given by:
\begin{equation}
    \centering
    \label{SYK-H}
    H = \sum_{ijkl}J_{ij;kl}c_i^\dagger c_j^\dagger c_kc_l
\end{equation}
where $i,j,k,l=1,2,...,L$ ($L$ is the system size). $J_{ij;kl}$ is a complex matrix whose independent elements follow normal distributions with:
\begin{equation}
    \centering
    \label{SYK-H-normal}
        \overline{J_{ij;kl}} = 0,\;\;\;\;\overline{{\left|J_{ij;kl}\right|}^2}=\frac{6J^2}{L^3} 
\end{equation}
where the overline denotes average over random realizations. It is worth recalling that the Hilbert space of real (Majorana) SYK with $L$ sites scales as $2^{L/2}$, while in the complex case it scales as $2^L$. However, the advantage of using this version of SYK \cite{Sonner:2017hxc} lies in the fact that the total number operator $\widehat{n}:=\sum_{i=1}^{L}c_i^\dagger c_i$ commutes with the Hamiltonian, allowing us to work in sectors of the Hilbert space with fixed occupation number, denoted by $N$ (eigenvalue of $\widehat{n}$). For our numerical computations we will take $N=\big{\lceil}{\frac{L}{2}}\big{\rceil}$.

The (hermitian) observable chosen for numerical computations is the hopping operator between sites $L$ and $L-1$:
\begin{equation}
    \centering
    \label{SYK-Hopping}
    \mathcal{O} = c_{L-1}^\dagger c_L+c_{L}^\dagger c_{L-1}\equiv h_{L-1,L} ~.
\end{equation}
Any other hopping operator $h_{ij}$ should give results equivalent to (\ref{SYK-Hopping}), due to the non-local character of the Hamiltonian (\ref{SYK-H}). In general, one could choose other non-extensive operators, such as the number operator at a particular site $n_i:=c_i^\dagger c_i$. Both $n_i$ and $h_{ij}$ have been shown numerically to satisfy ETH \cite{Sonner:2017hxc}. 

\begin{figure*}[t]
    \begin{minipage}{.5\textwidth}
    \includegraphics[width=1.\linewidth]{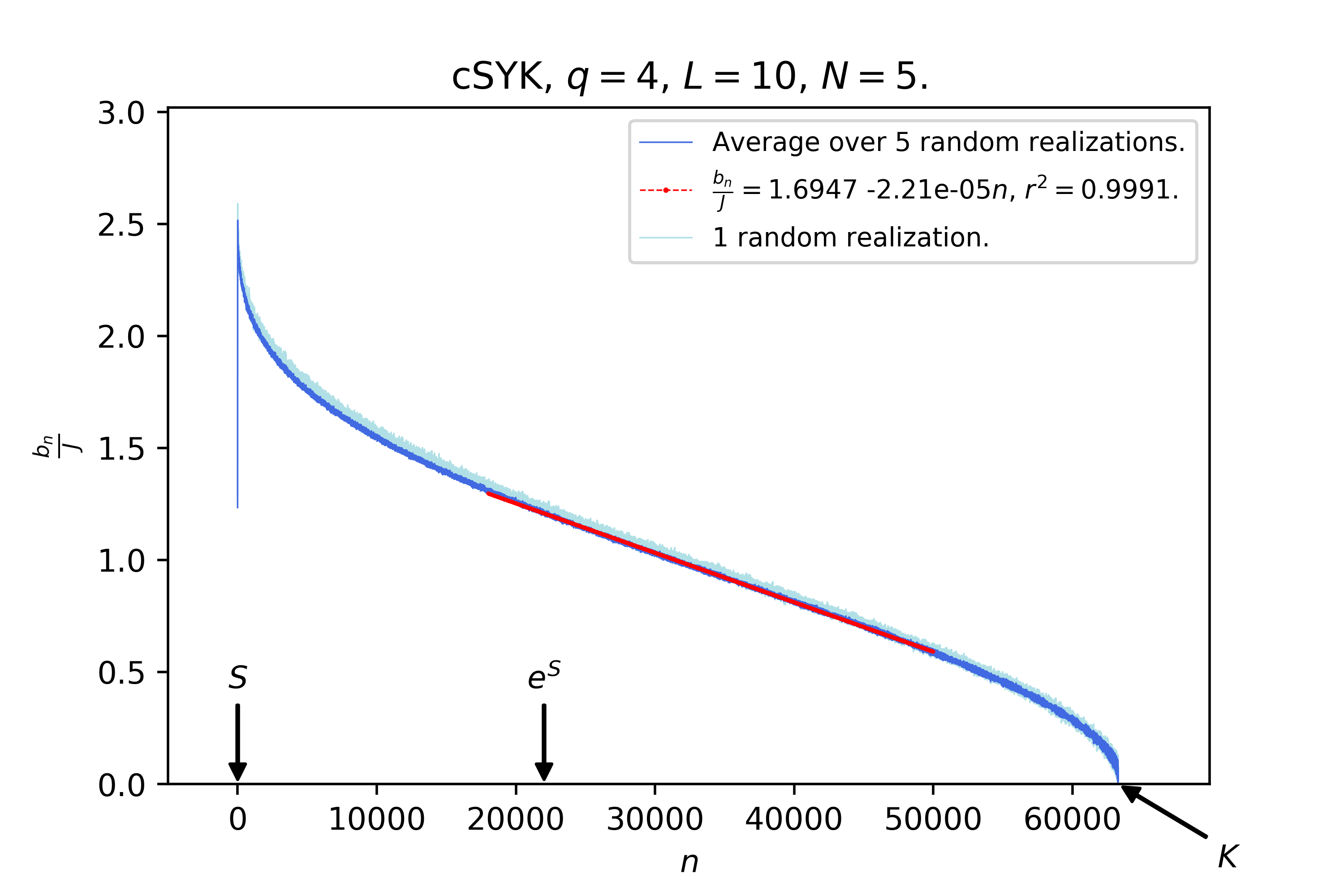}
    \end{minipage}  \quad  
    \begin{minipage}{.5\textwidth}
    \includegraphics[width=1.\linewidth]{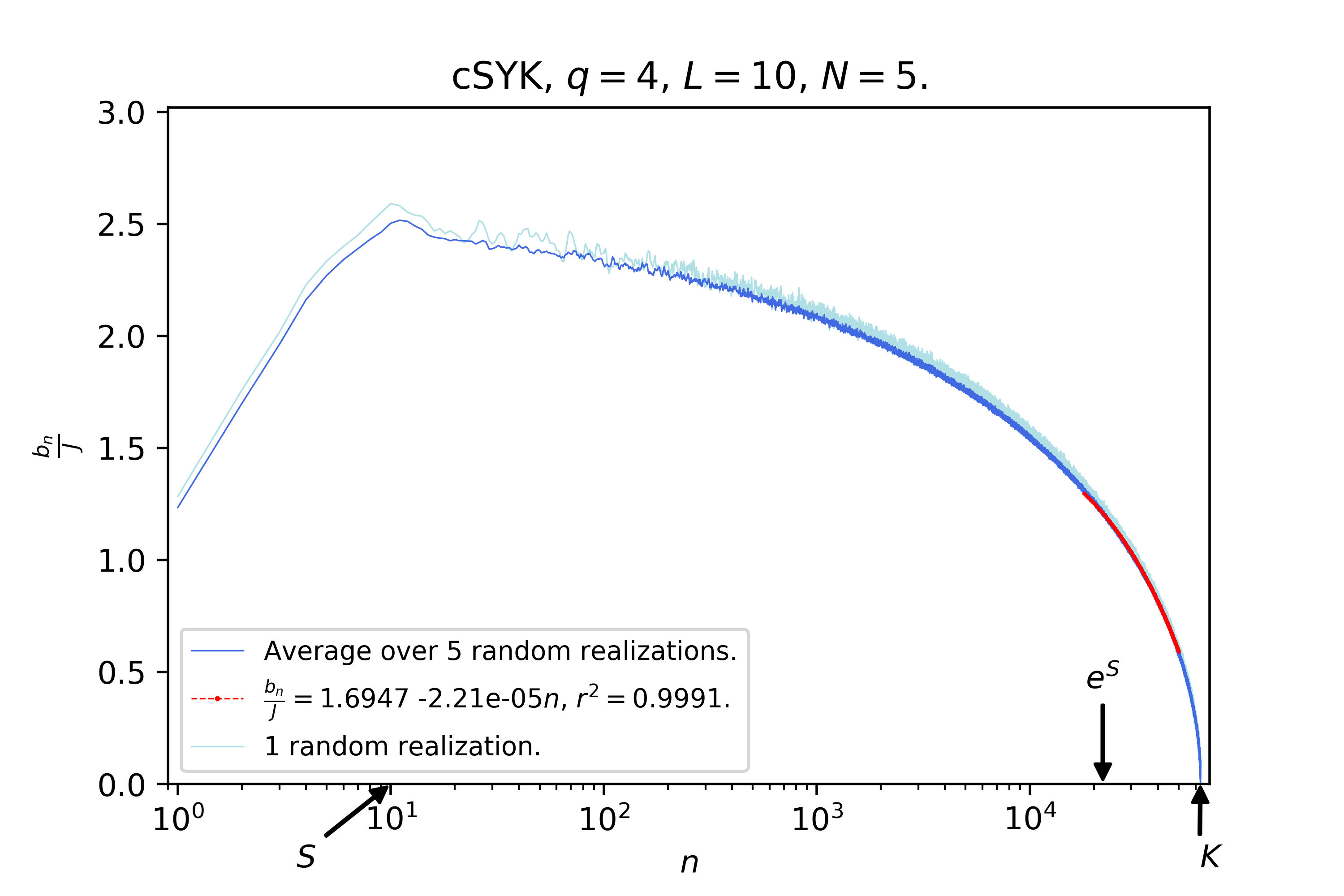}
    \end{minipage} 
    \caption{Lanczos-sequence for $L=10$. \textbf{Left:} ``The Descent" depicted with linear scale along the horizontal axis. After initial growth up to $n\sim S$, a slow decrease to zero with roughly constant negative non-perturbative slope of order $\sim -\frac{1}{K} \sim - e^{-2S}$. Fitted slope (in red) of the decaying part is $-2.21\cdot 10^{-5}\approx-1.58\cdot10^{-5}=-K^{-1}$. On a 1:1 scale, the horizontal axis should be at least $843$ meters long.  \textbf{Right:} Logarithmic scale along the horizontal axis makes visible the initial ascent.}
    \label{b-L10-main}
\end{figure*}

\begin{figure*}
    \begin{minipage}{.5\textwidth}
	\includegraphics[width=1.\linewidth]{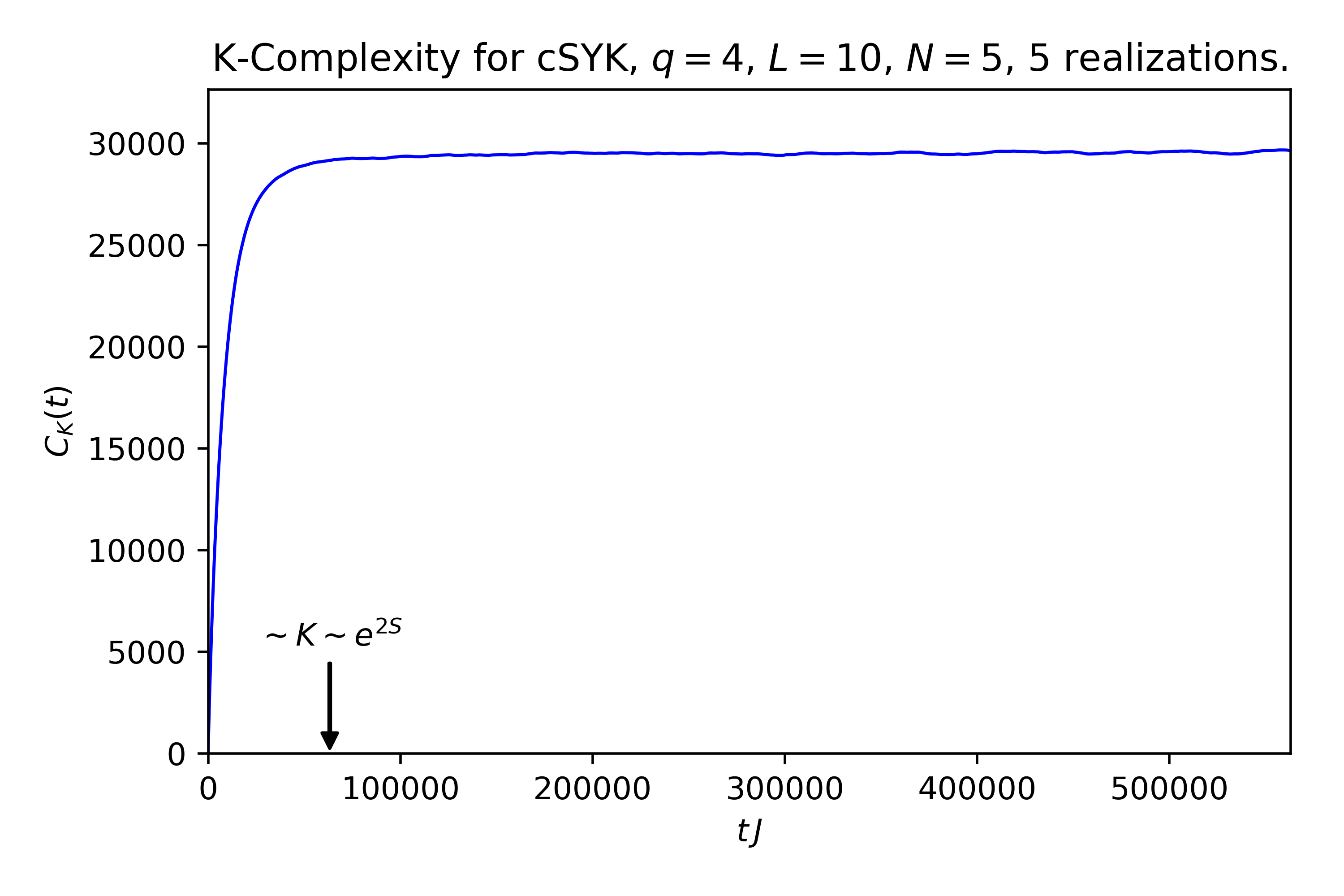}
	\end{minipage}  \quad 
	\begin{minipage}{.5\textwidth}
	\includegraphics[width=1.\linewidth]{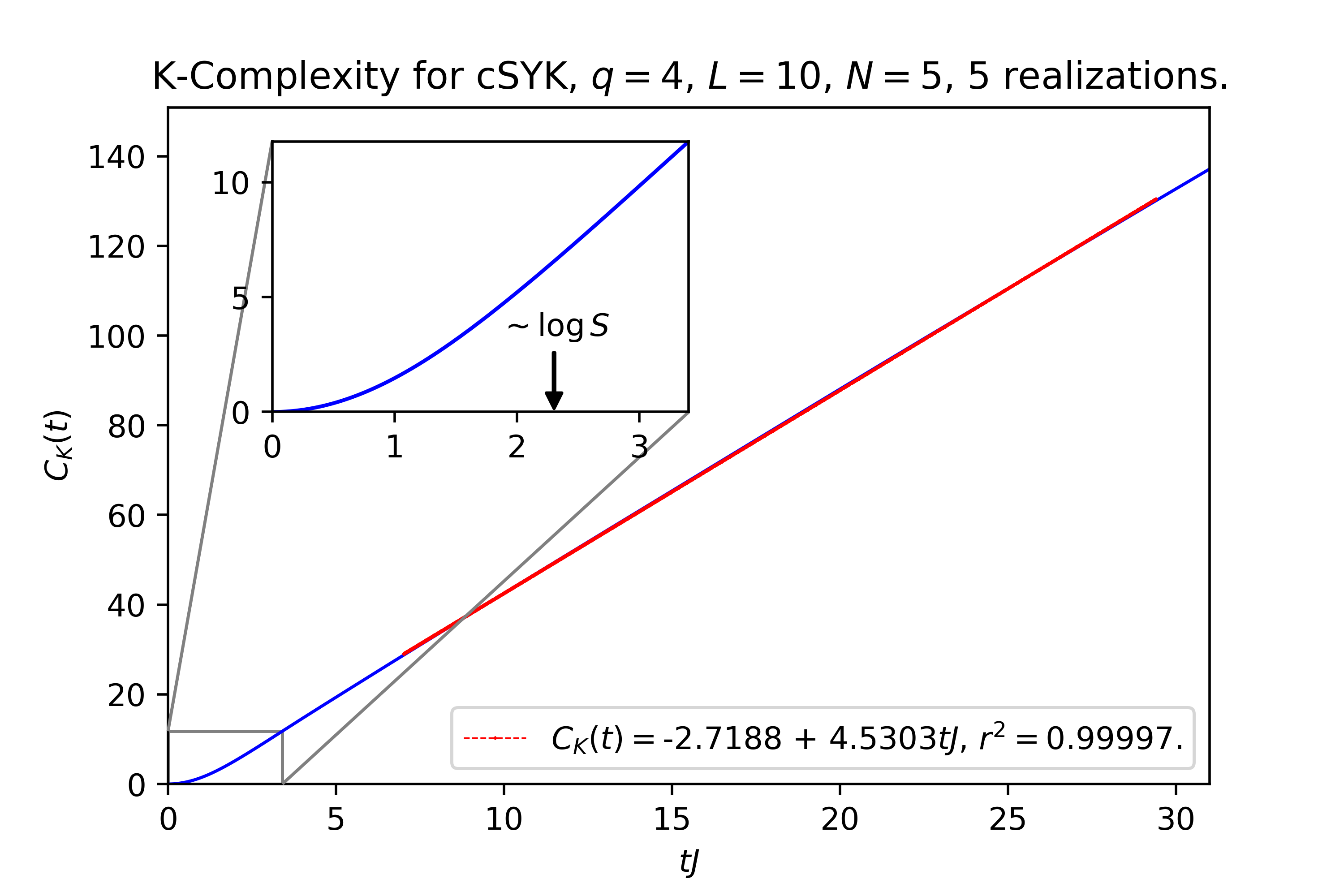}
	\end{minipage}
	\caption{K-complexity averaged over 5 random realizations with $L=10$ sites at half filling. \textbf{Left:} Full time range computed. Saturation occurs at time scales of order $t_K\sim K \sim e^{2S}$ with value near $K/2=31626.5$ (this is the time scale at which the wave-packet $\varphi_n(t)$, which propagates at roughly constant -- but slowly decreasing -- velocity, reaches the edge of the Krylov chain).  \textbf{Right:} Zoom in at early times. Note the initial non-linear growth transforming into linear growth, as verified by the linear fit (in red). Relevant time scales are indicated in the plots, although their exact location may depend on a dimensionful prefactor (the Lyapunov exponent, as discussed in \cite{Barbon:2019wsy}).}  
	\label{KC-L10-main}
\end{figure*}

\begin{figure*}
    \begin{minipage}{.5\textwidth}
	\includegraphics[width=1.\linewidth]{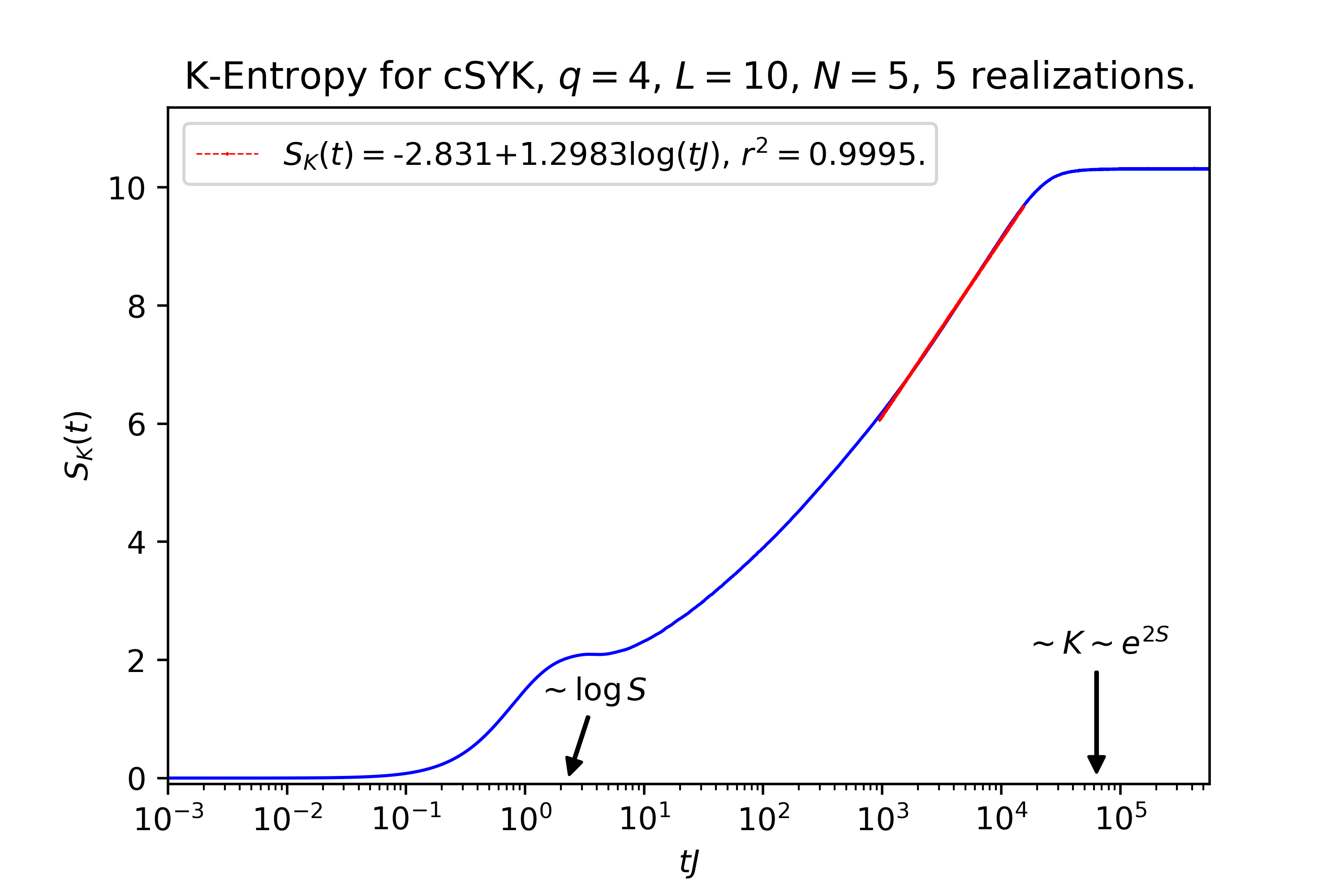}
    \end{minipage}  \quad 
    \begin{minipage}{.5\textwidth}
    \includegraphics[width=1.\linewidth]{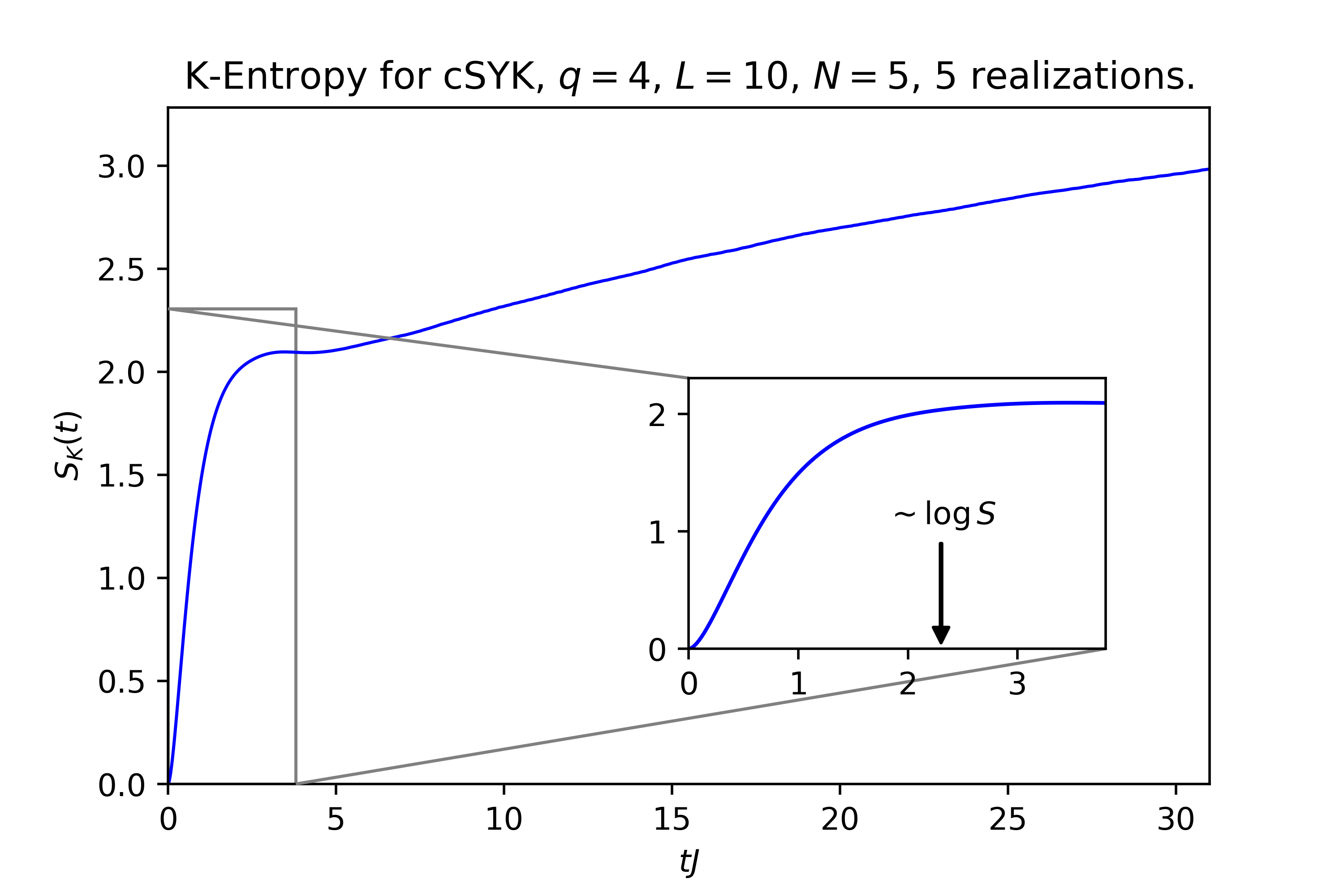}
    \end{minipage}
    \caption{K-entropy averaged over 5 random realizations with $L=10$ sites at half filling. \textbf{Left:} Full time range computed, with logarithmic scale along horizontal axis. A linear fit in the post-scrambling regime, where K-entropy is expected to grow logarithmically, is depicted in red. K-entropy grows linearly up to scrambling time, and then transitions to a logarithmic growth phase that continues until saturation around $S_K\sim L\sim S$ at exponentially late times (this is the time scale at which the wave-packet $\varphi_n(t)$ becomes fully dispersed). \textbf{Right:} K-entropy at early times with linear scale along the horizontal axis. The exact location of the time scales may depend on a dimensionful prefactor (see caption of Figure \ref{KC-L10-main}). 
}
    \label{KS-L10-main}
\end{figure*}

\begin{figure*}[t]
\centering
\begin{minipage}{.35\textwidth}
    \includegraphics[width=1.\linewidth]{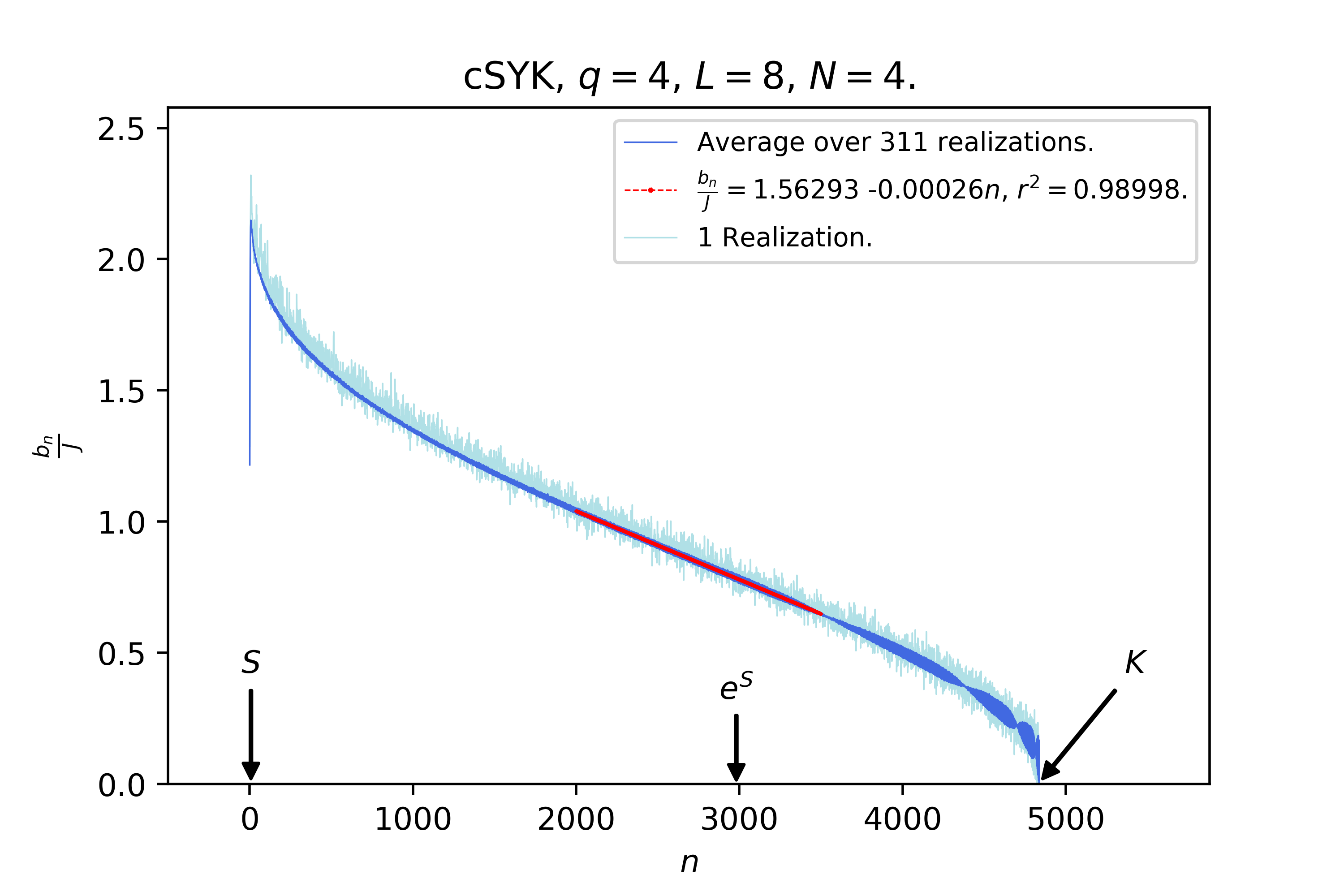}
\end{minipage} \qquad
\begin{minipage}{.35\textwidth}
    \includegraphics[width=1.\linewidth]{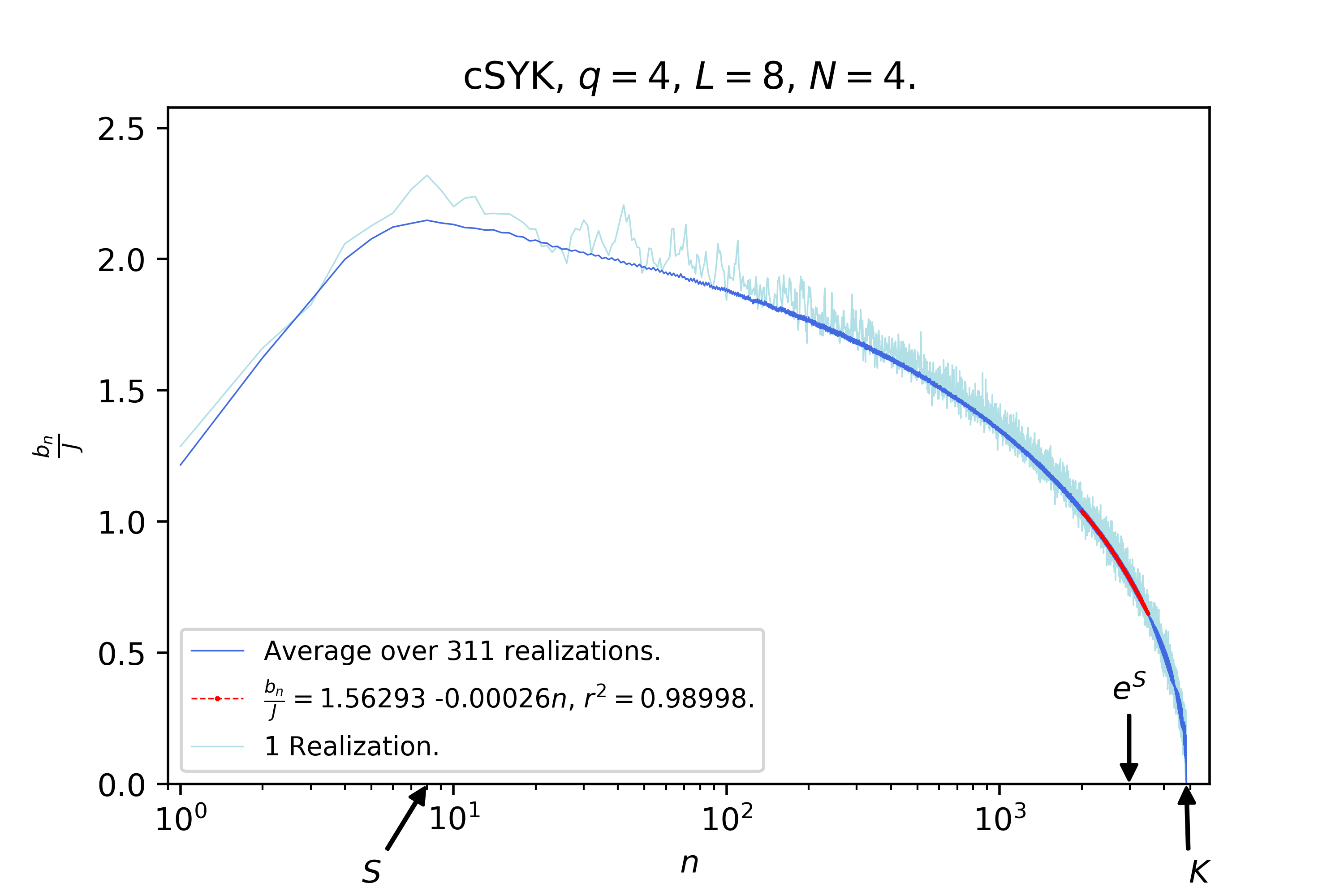}
\end{minipage} \\
\begin{minipage}{.35\textwidth}
\includegraphics[width=1.\linewidth]{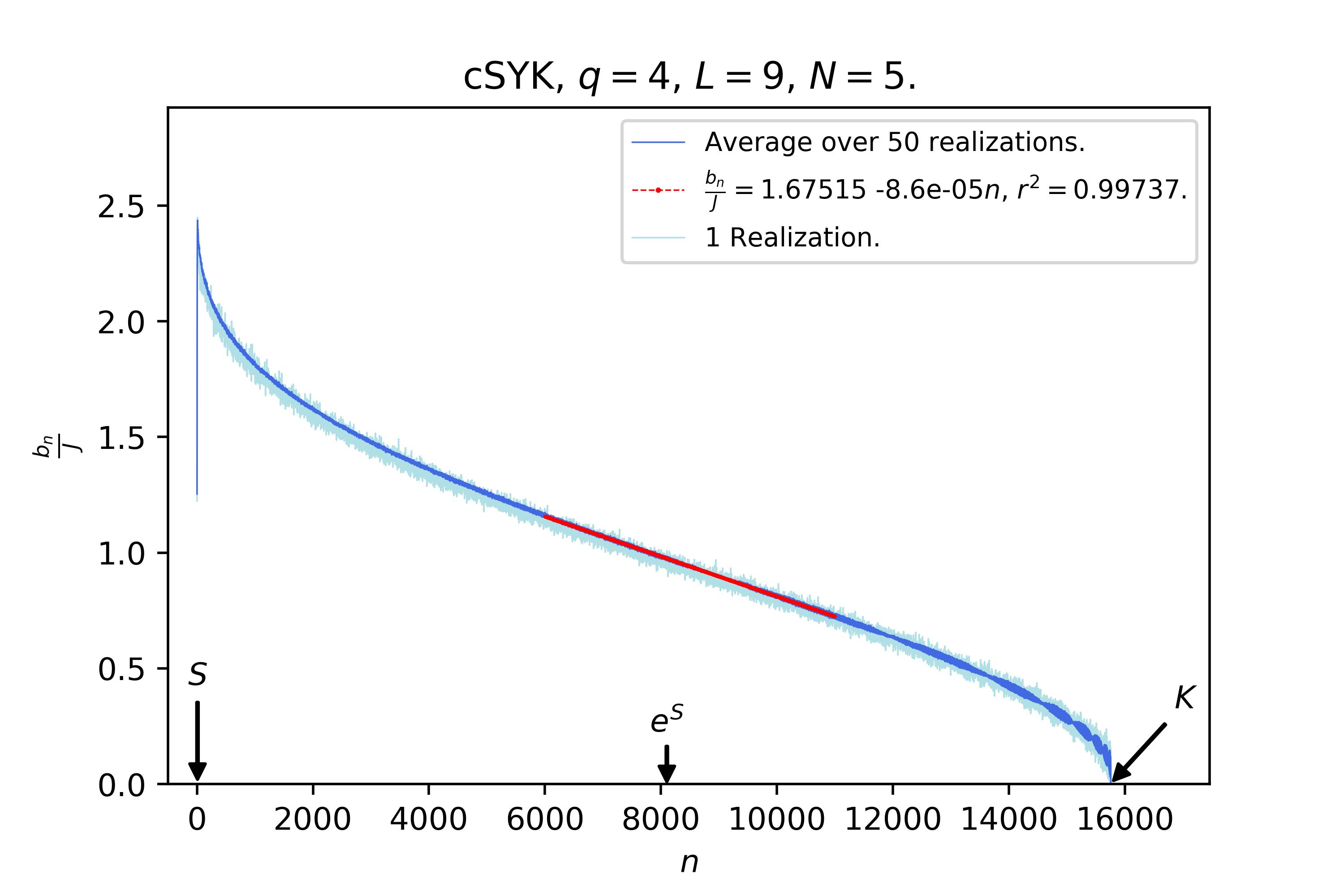}
\end{minipage}\qquad
\begin{minipage}{.35\textwidth}
\includegraphics[width=1.\linewidth]{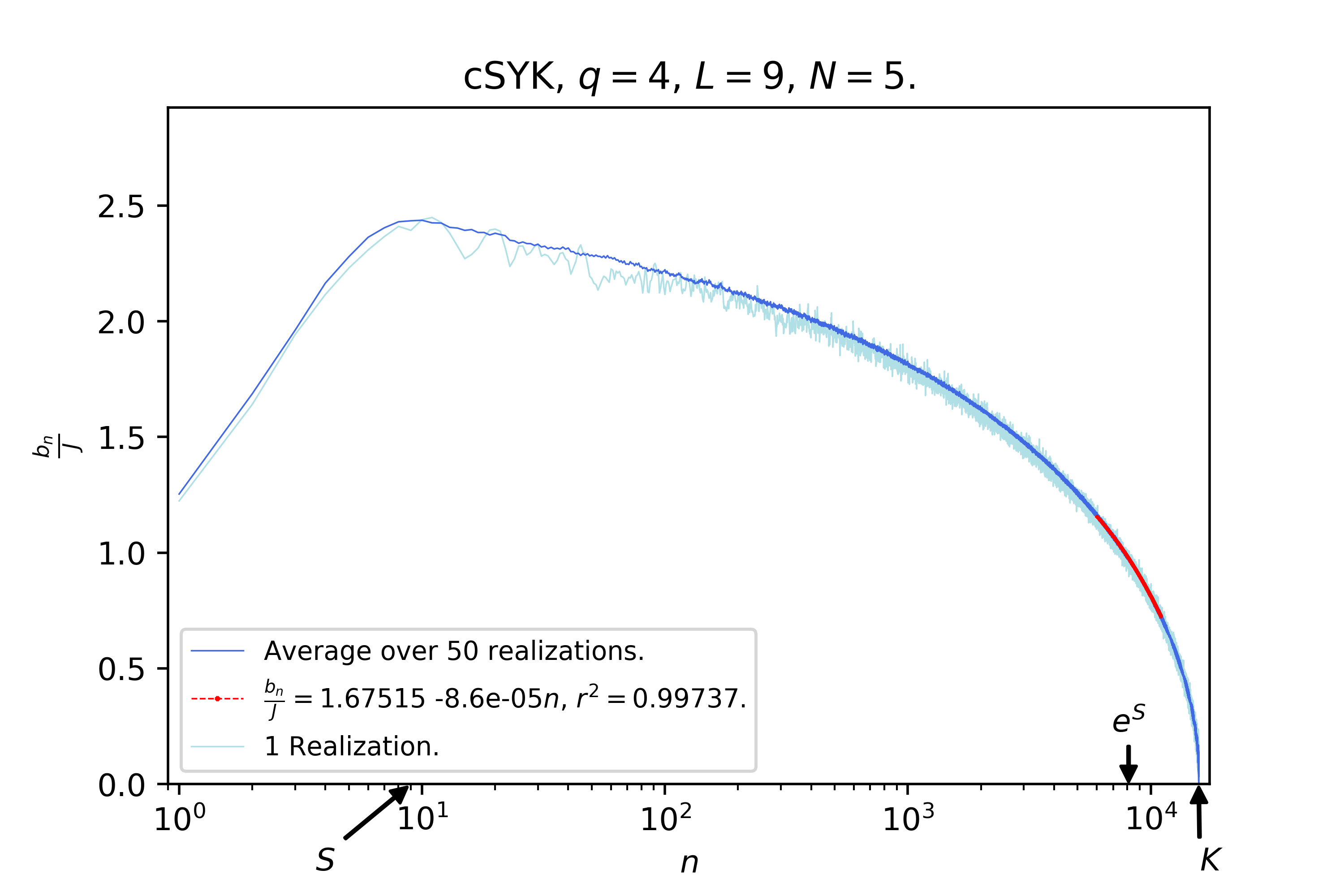}
\end{minipage} \\
\begin{minipage}{.35\textwidth}
\includegraphics[width=1.\linewidth]{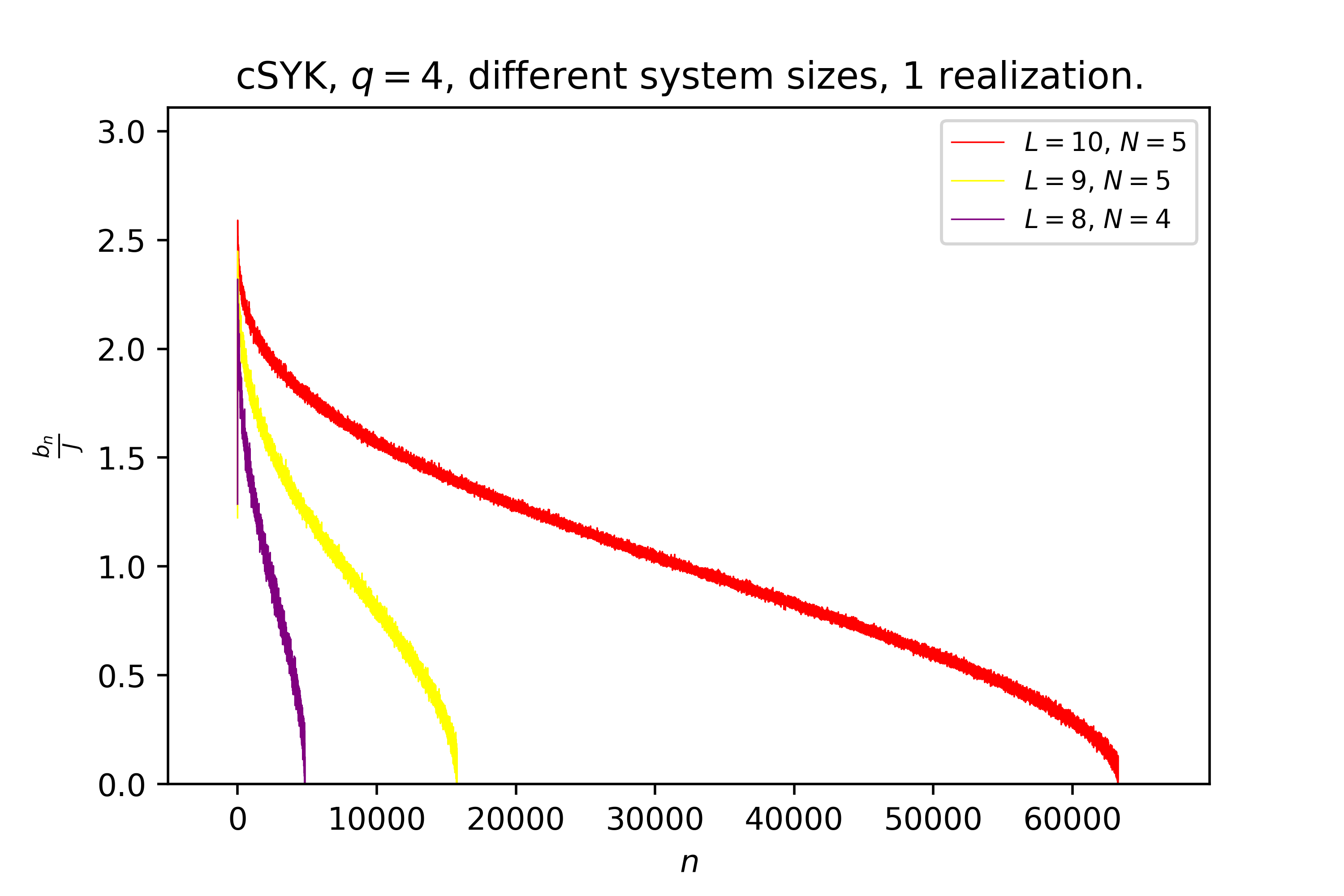}
\end{minipage}\qquad
\begin{minipage}{.35\textwidth}
\includegraphics[width=1.\linewidth]{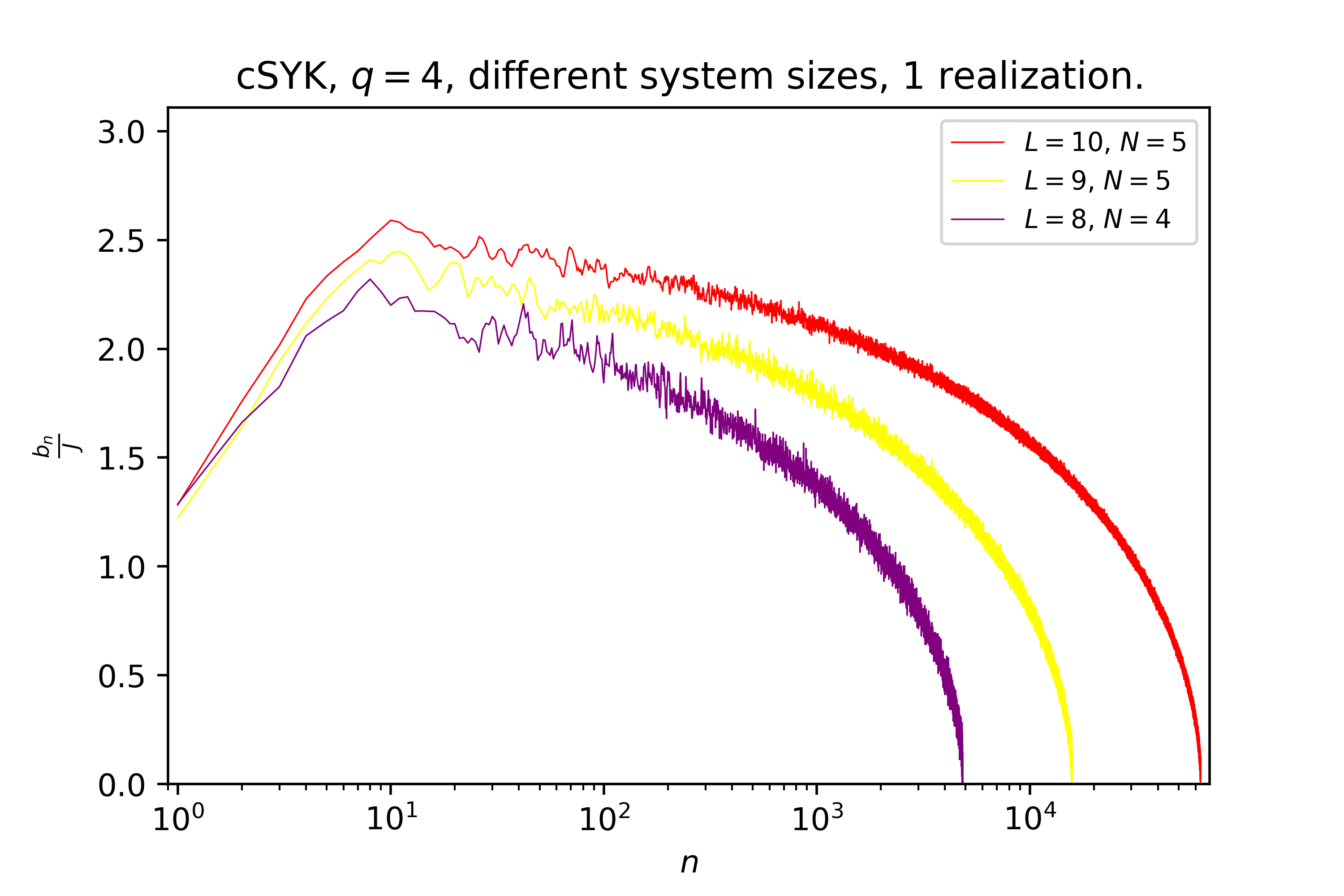}
\end{minipage}
\caption{Lanczos sequences for $L=8,\, N=4$ and $L=9,\, N=5$ and comparison of results for $L=8,9,10$. \textbf{Top row:} Results for $L=8$ in linear  (left panel) and logarithmic (right panel) scale along the horizontal axis. The plots depict both the sequence of a single random realization and the average over $311$ realizations. A linear fit is included in the decaying tail, whose slope approaches numerically the na\"ive estimate $\sim - \frac{1}{K}\approx - 0.000206$.  \textbf{Middle row:} Results for $L=9$ in linear  (left panel) and logarithmic (right panel) scale along the horizontal axis. The plots depict both the sequence of a single random realization and the average over $50$ realizations. A linear fit is included in the decaying tail, whose slope is of the order of the naive estimate $\sim-\frac{1}{K}\approx -6.3\cdot10^{-5}$.  \textbf{Bottom row:} Comparison of the Lanczos sequences for $L=8,9,10$ in linear (left panel) and logarithmic (right panel) scale along the horizontal axis.}
\label{b-sequences_8_9_Comp}
\end{figure*}

\begin{figure*}[t]
\centering
\begin{minipage}{.35\textwidth}
    \includegraphics[width=1.\linewidth]{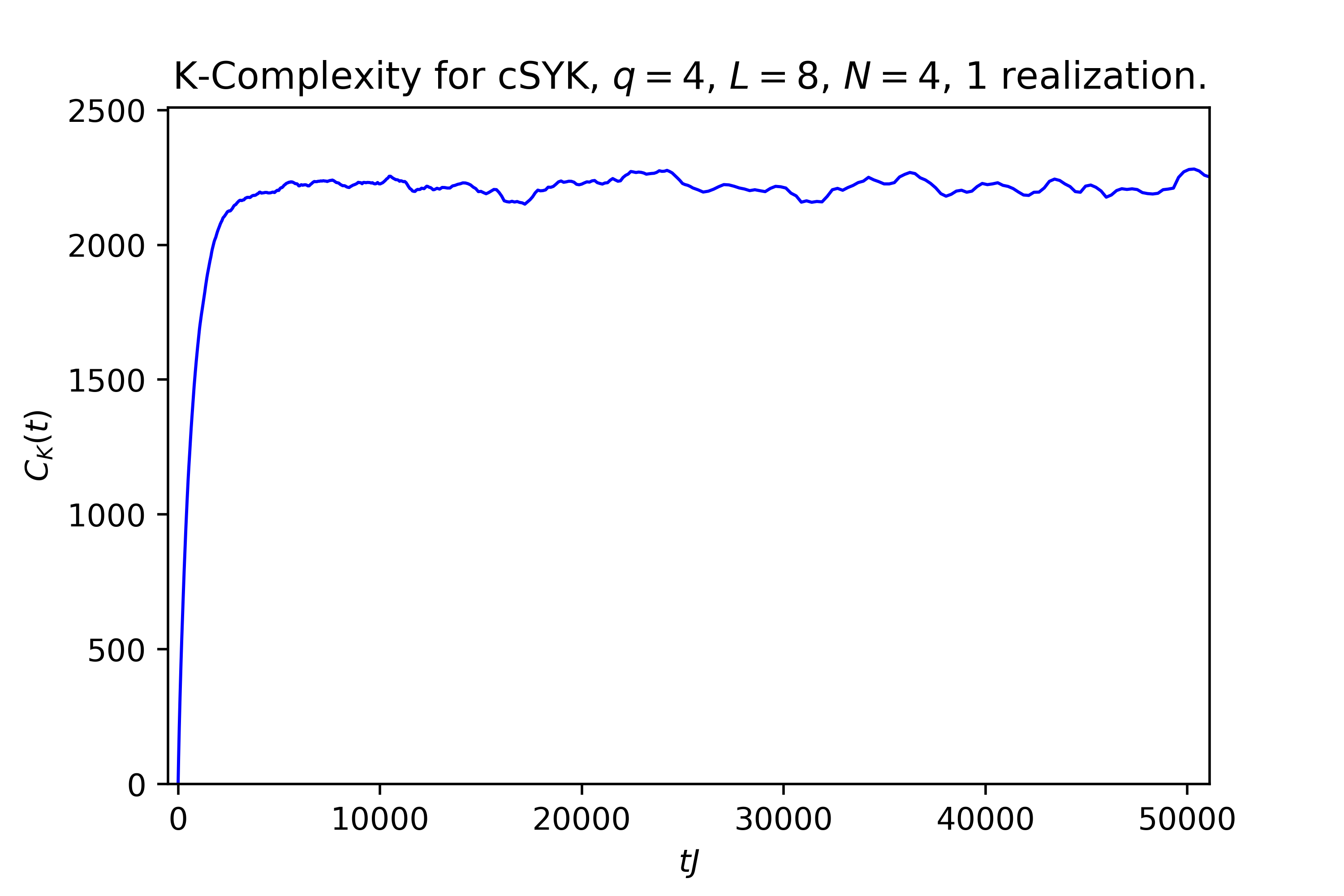}
\end{minipage} \qquad
\begin{minipage}{.35\textwidth}
    \includegraphics[width=1.\linewidth]{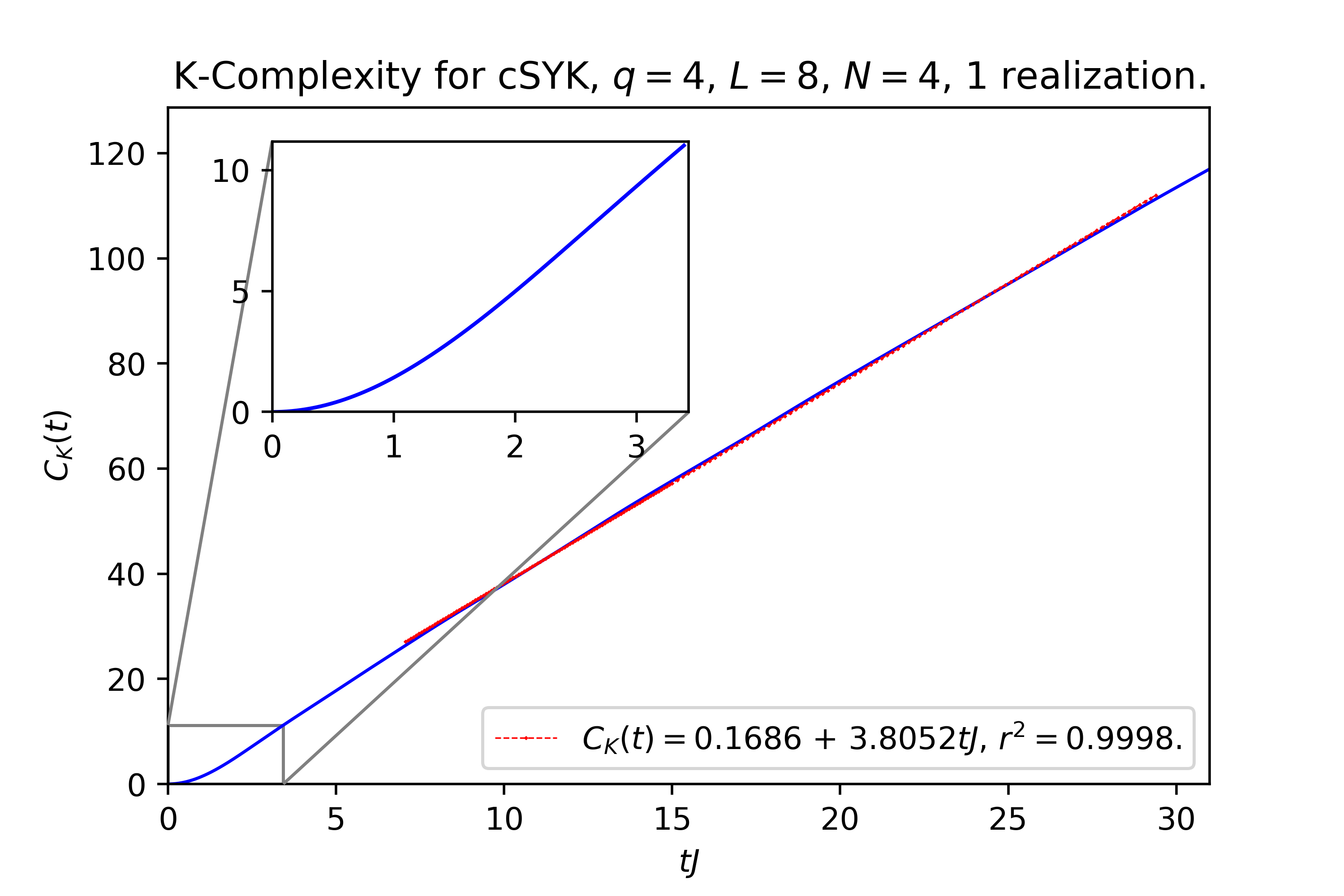}
\end{minipage} \\
\begin{minipage}{.35\textwidth}
\includegraphics[width=1.\linewidth]{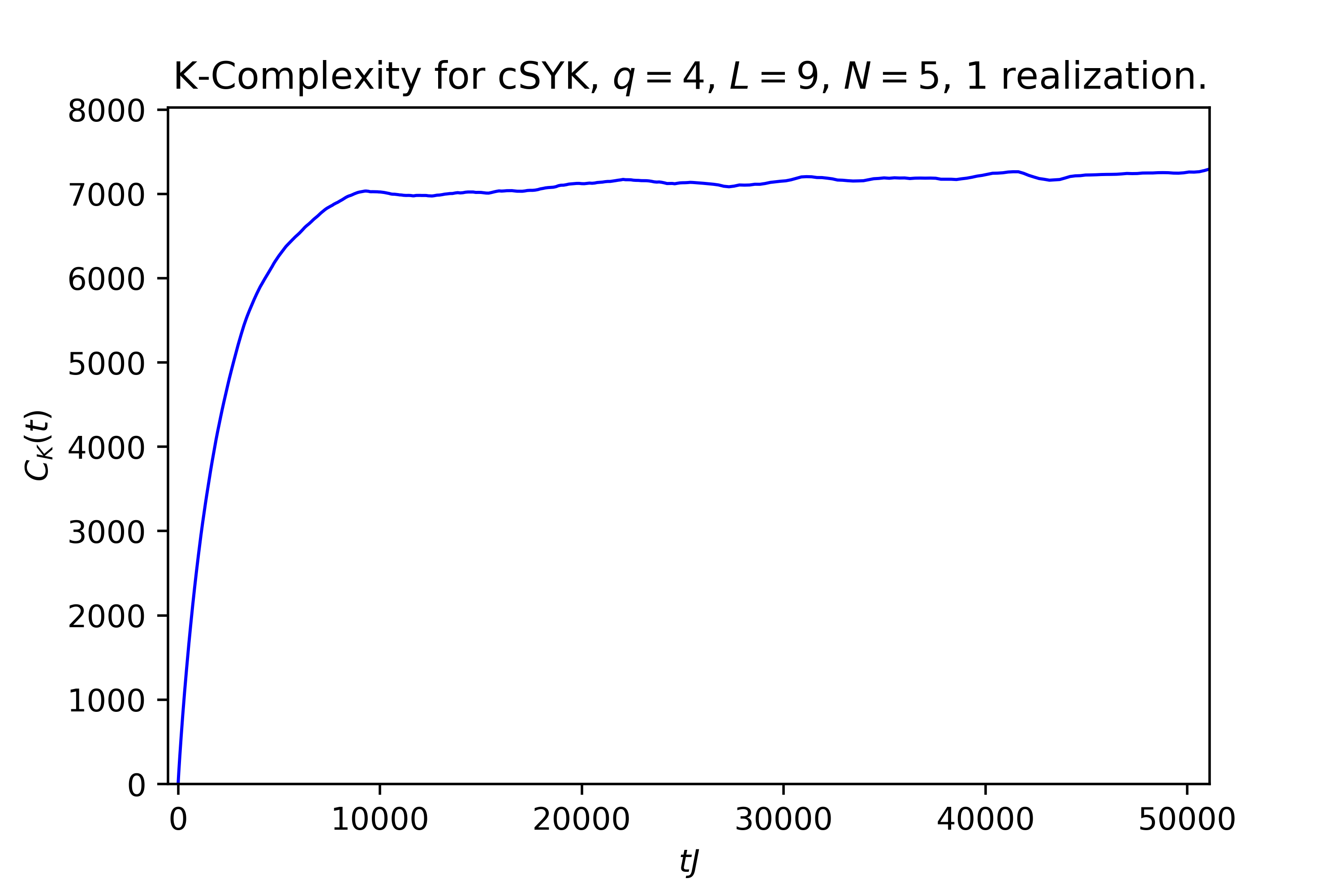}
\end{minipage}\qquad
\begin{minipage}{.35\textwidth}
\includegraphics[width=1.\linewidth]{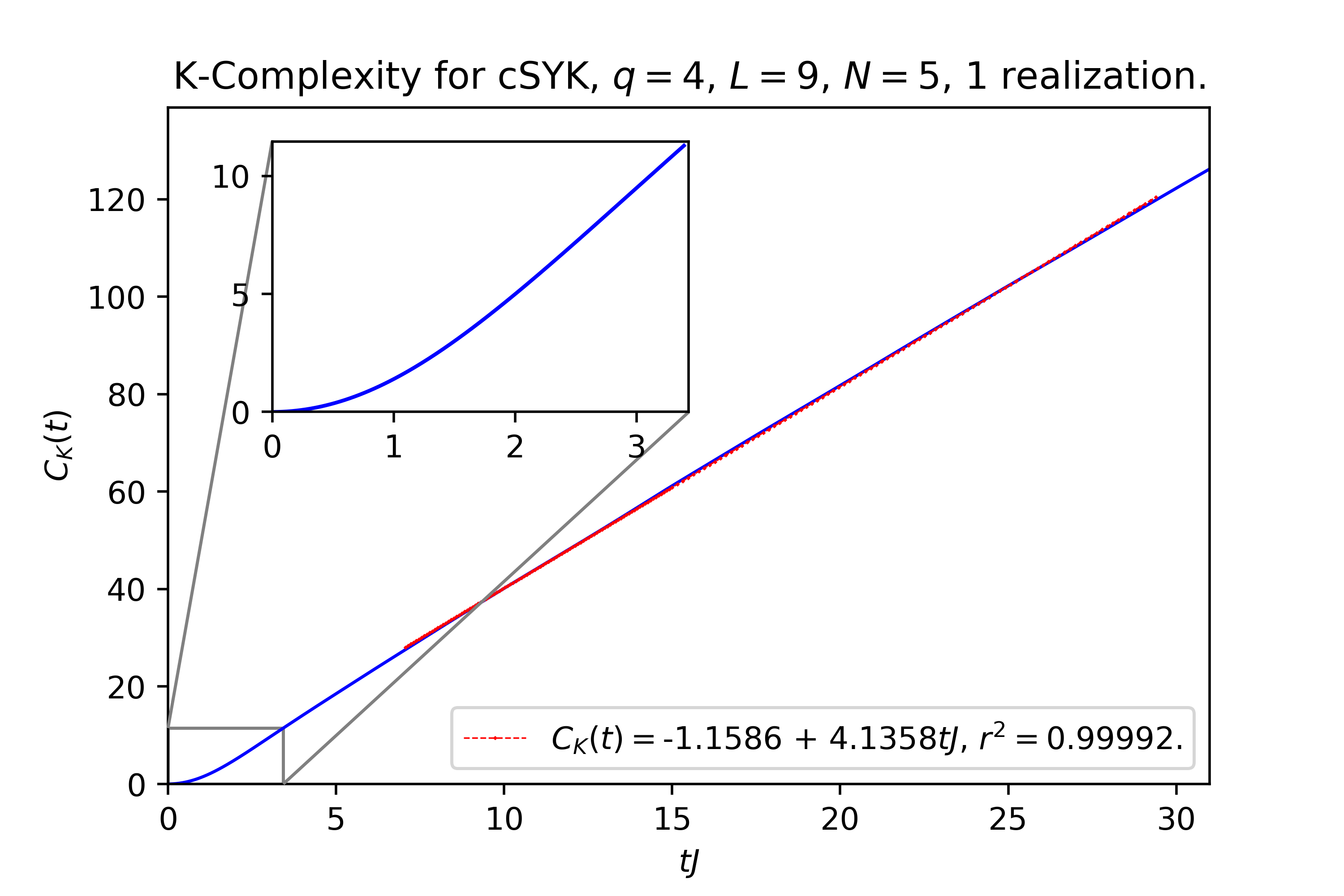}
\end{minipage} \\
\begin{minipage}{.35\textwidth}
\includegraphics[width=1.\linewidth]{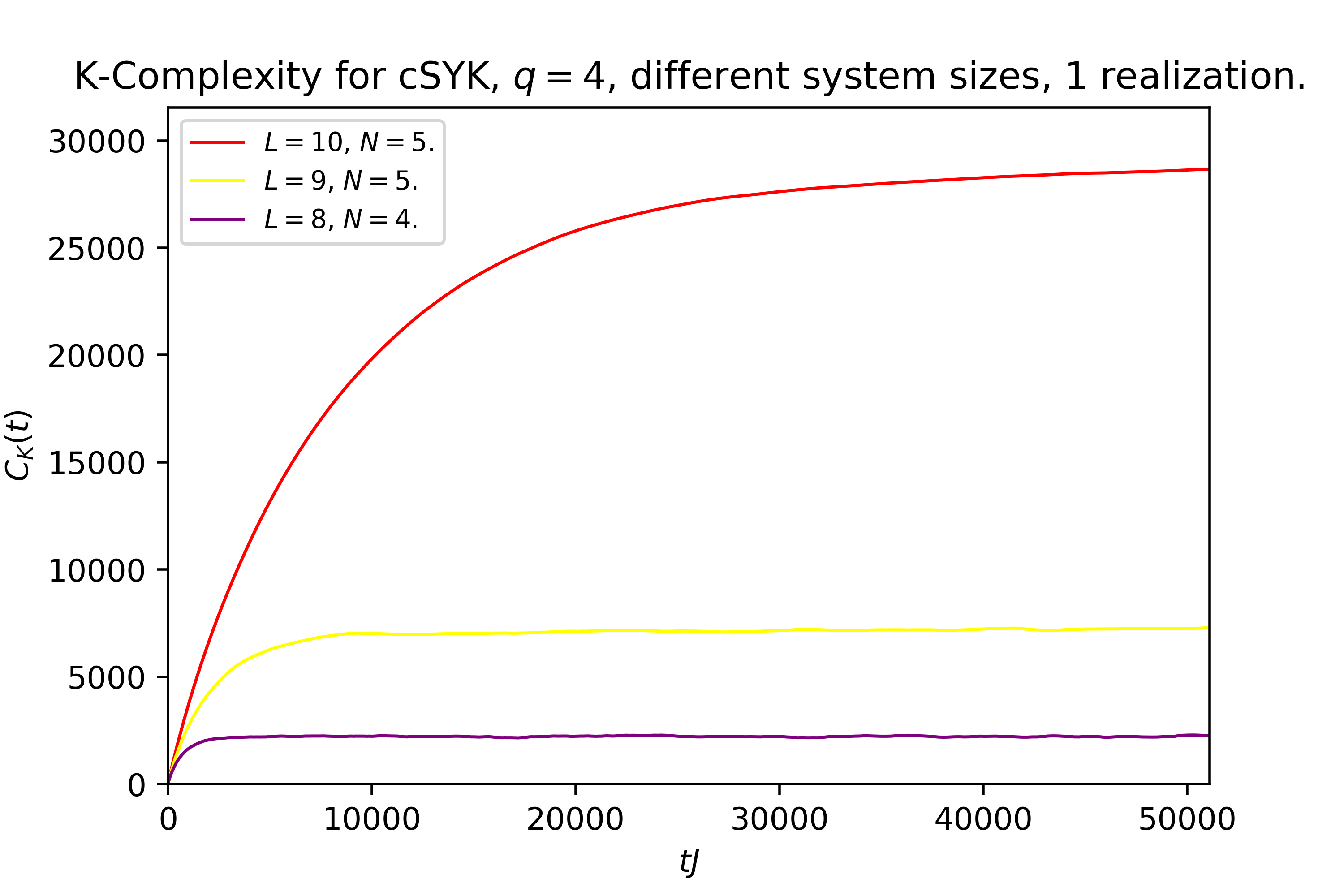}
\end{minipage}\qquad
\begin{minipage}{.35\textwidth}
\includegraphics[width=1.\linewidth]{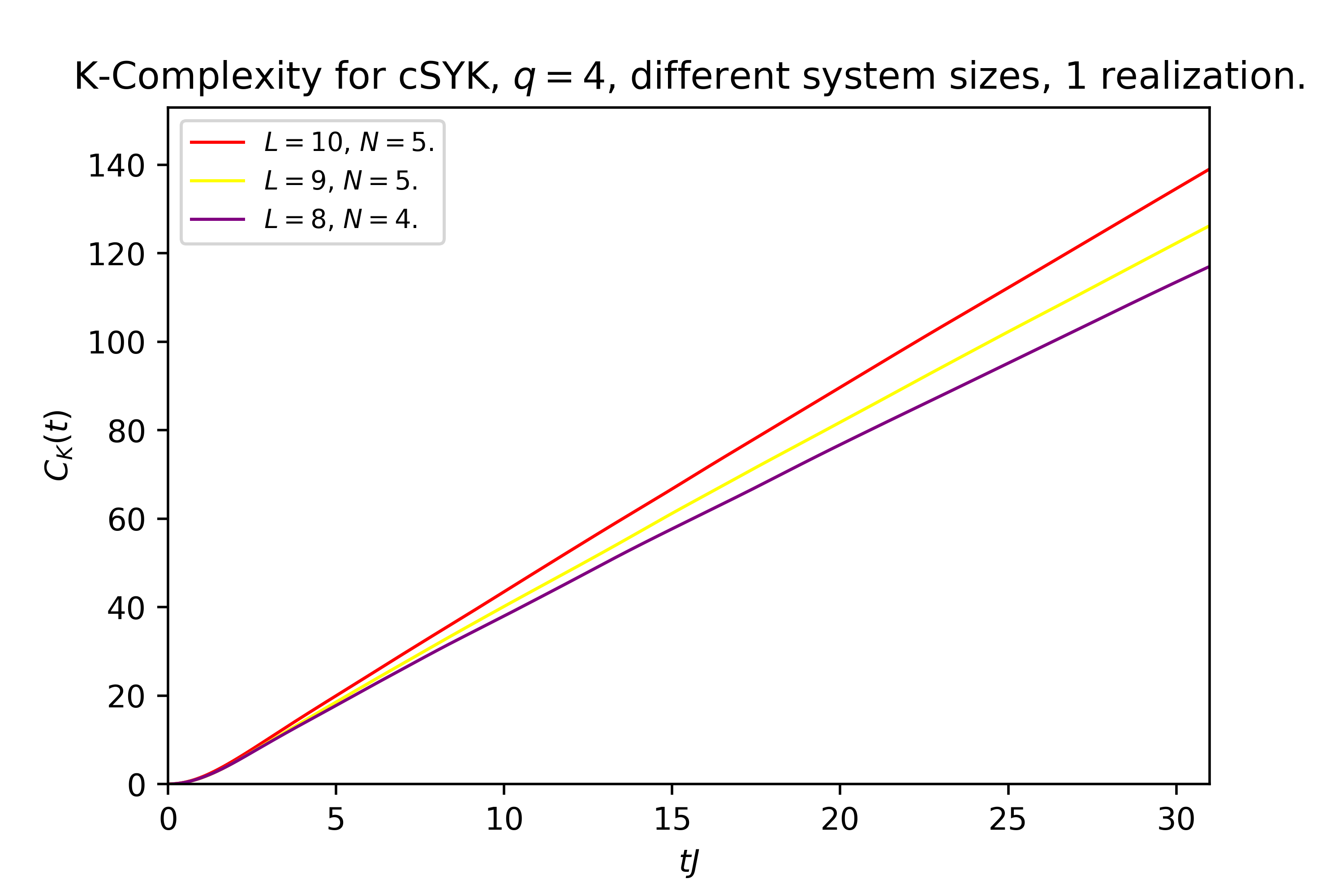}
\end{minipage}
\caption{Results for K-complexity for $L=8,\, N=4$ and $L=9,\, N=5$ and comparison of results for $L=8,9,10$. \textbf{Top row:} Results for $L=8$; for exponentially long times (left panel) and for early times (right panel). Note the change in behaviour from very early times (inset) and later times. The value at saturation is near $\sim\frac{K}{2}=2415.5$. \textbf{Middle row:} Results for $L=9$; for exponentially long times (left panel) and for early times (right panel). The value at saturation is near $\sim\frac{K}{2}=7875.5$. \textbf{Bottom row:} Comparison of results for $L=8,9,10$ for exponentially long times (left panel) and for early times (right panel).}
\label{KC_8_9_Comp}
\end{figure*}

\begin{figure*}[t]
\centering
\begin{minipage}{.3\textwidth}
    \includegraphics[width=1.\linewidth]{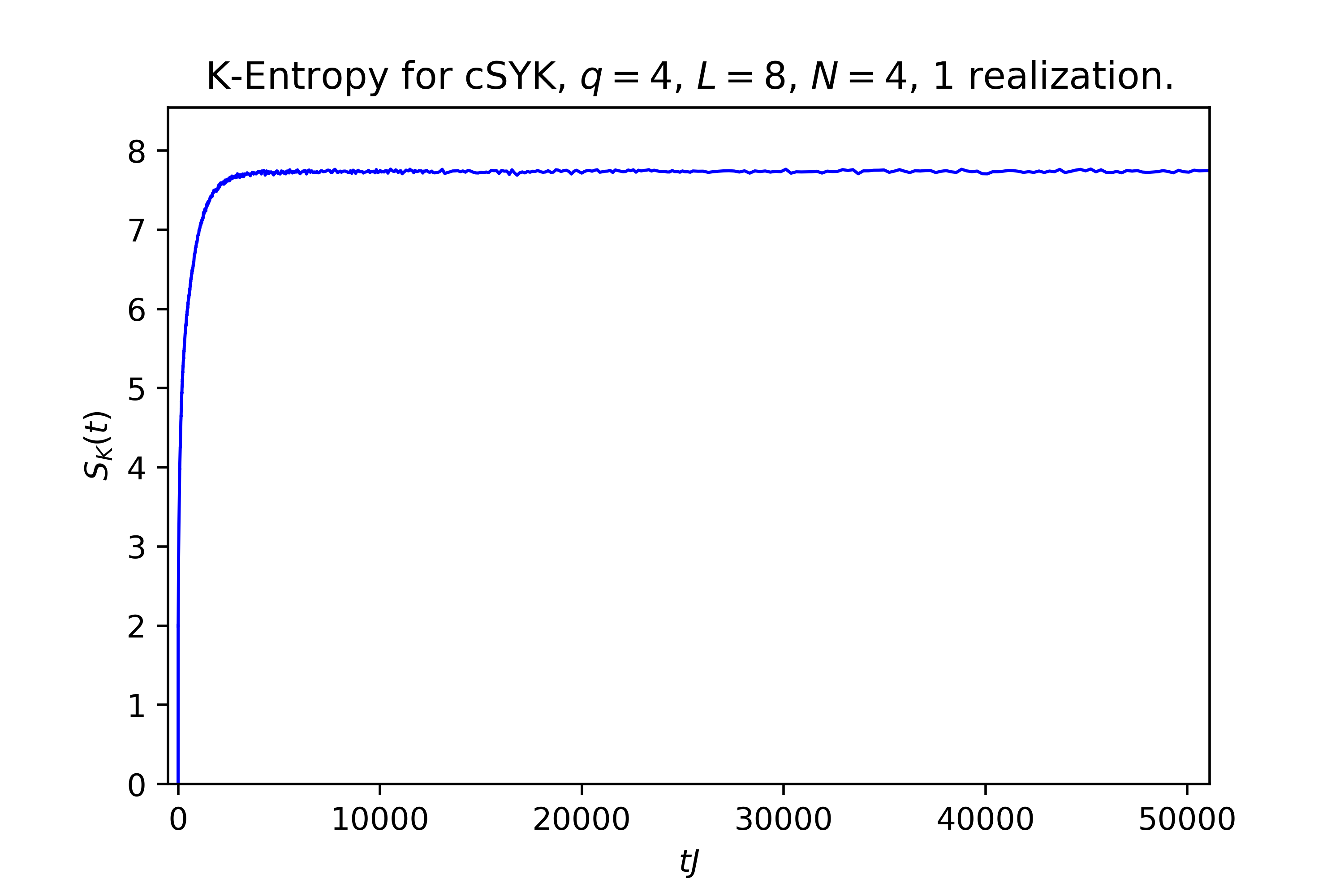}
\end{minipage} \quad
\begin{minipage}{.3\textwidth}
    \includegraphics[width=1.\linewidth]{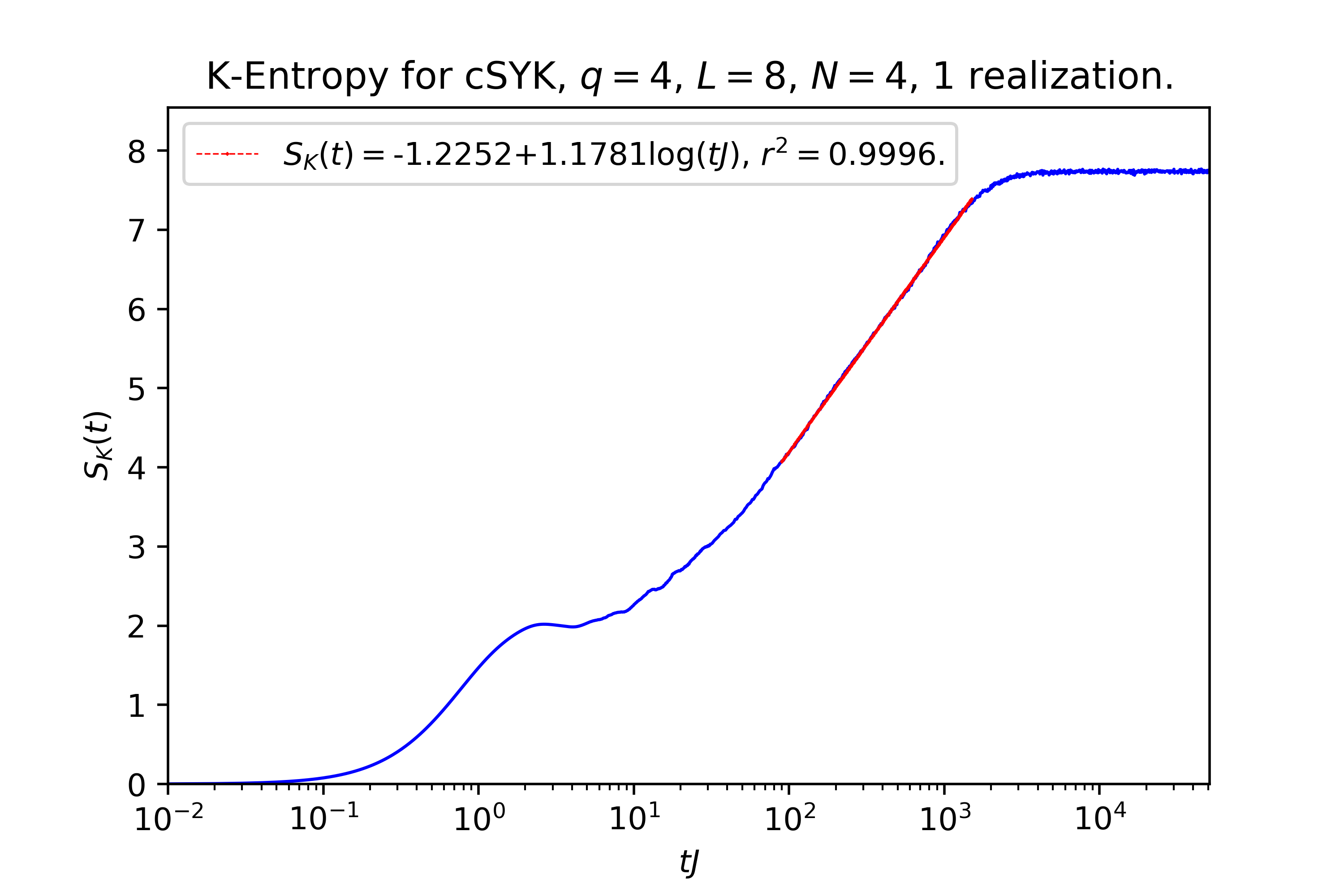}
\end{minipage} \quad
\begin{minipage}{.3\textwidth}
    \includegraphics[width=1.\linewidth]{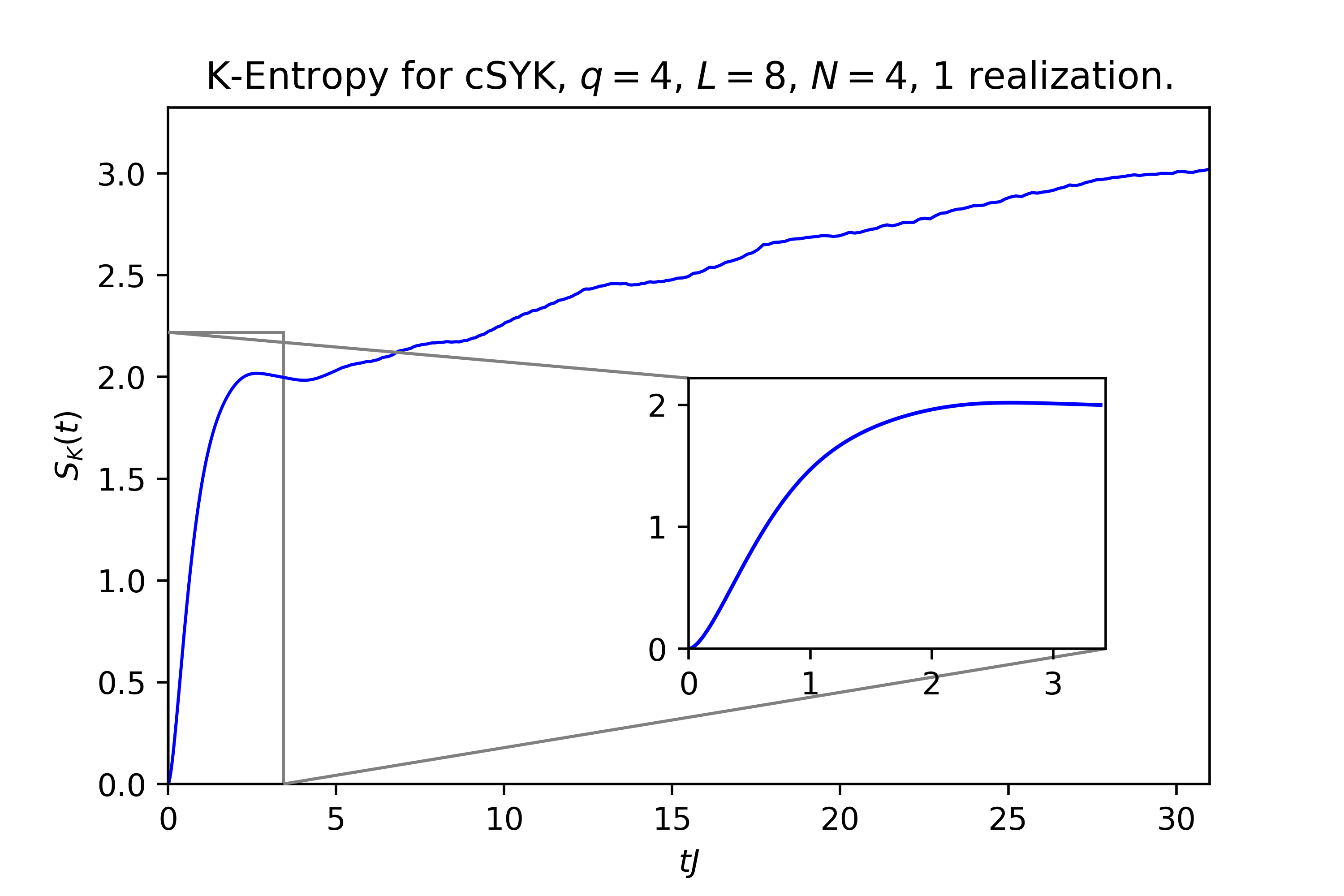}
\end{minipage}\\
\begin{minipage}{.3\textwidth}
    \includegraphics[width=1.\linewidth]{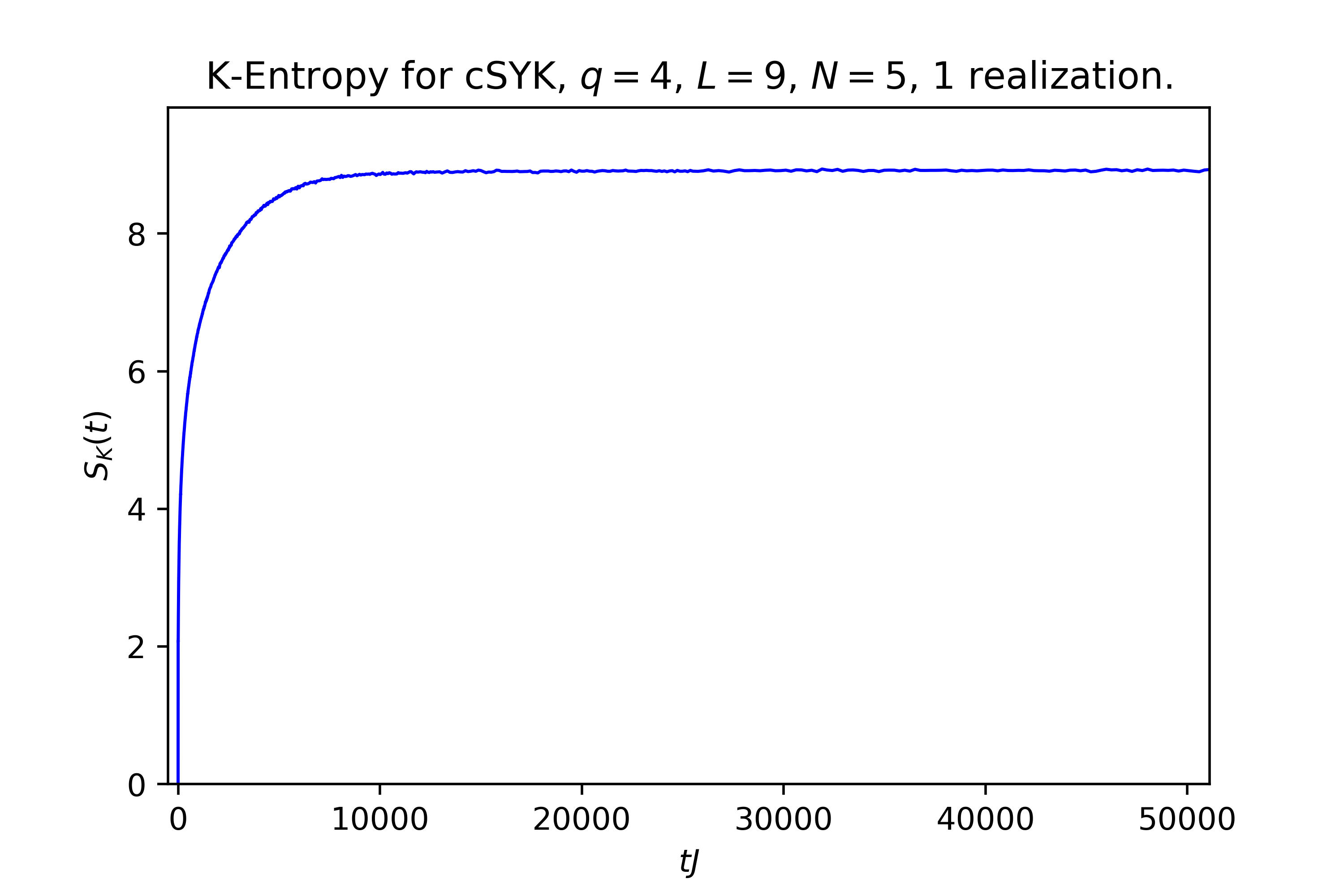}
\end{minipage} \quad
\begin{minipage}{.3\textwidth}
    \includegraphics[width=1.\linewidth]{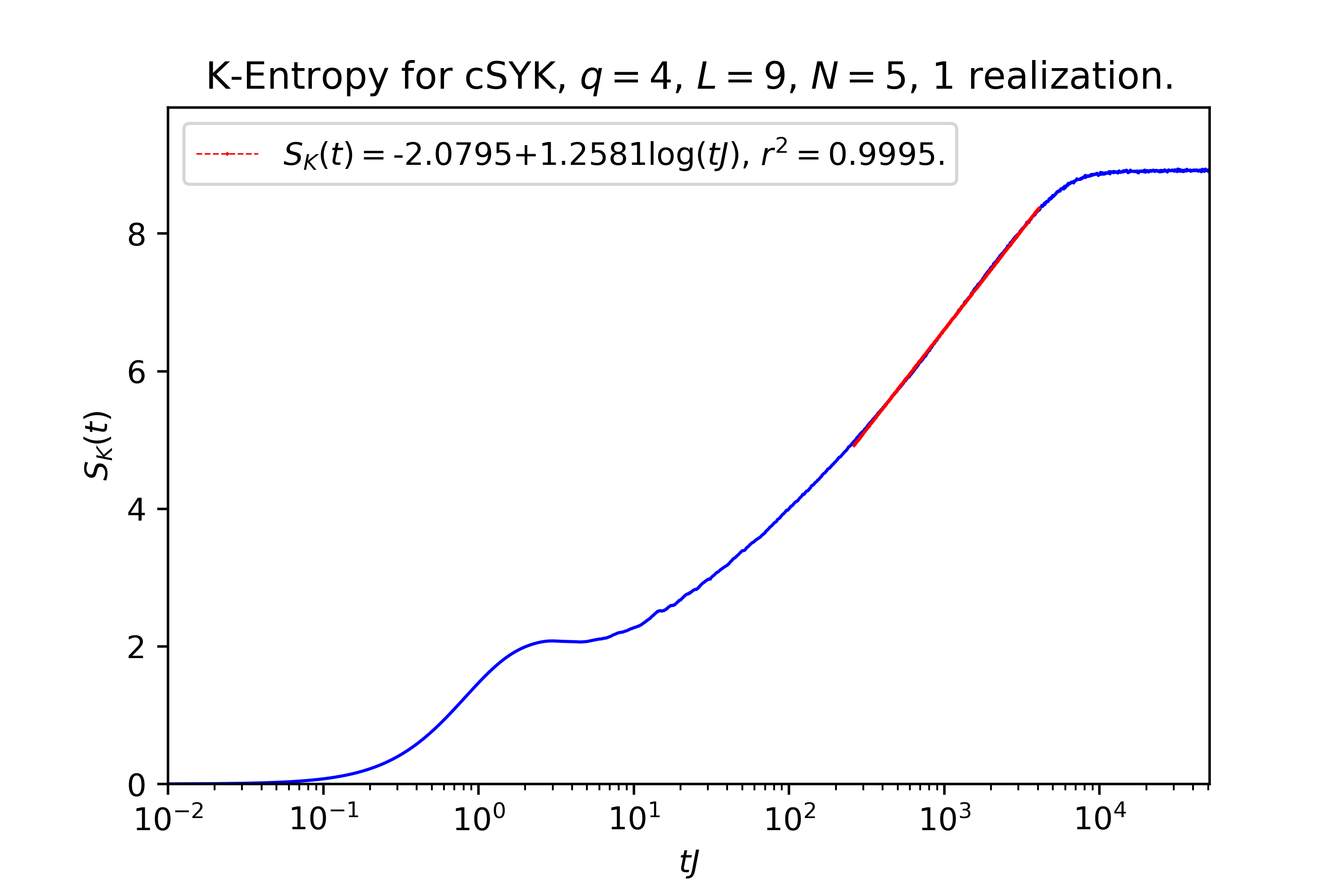}
\end{minipage} \quad
\begin{minipage}{.3\textwidth}
    \includegraphics[width=1.\linewidth]{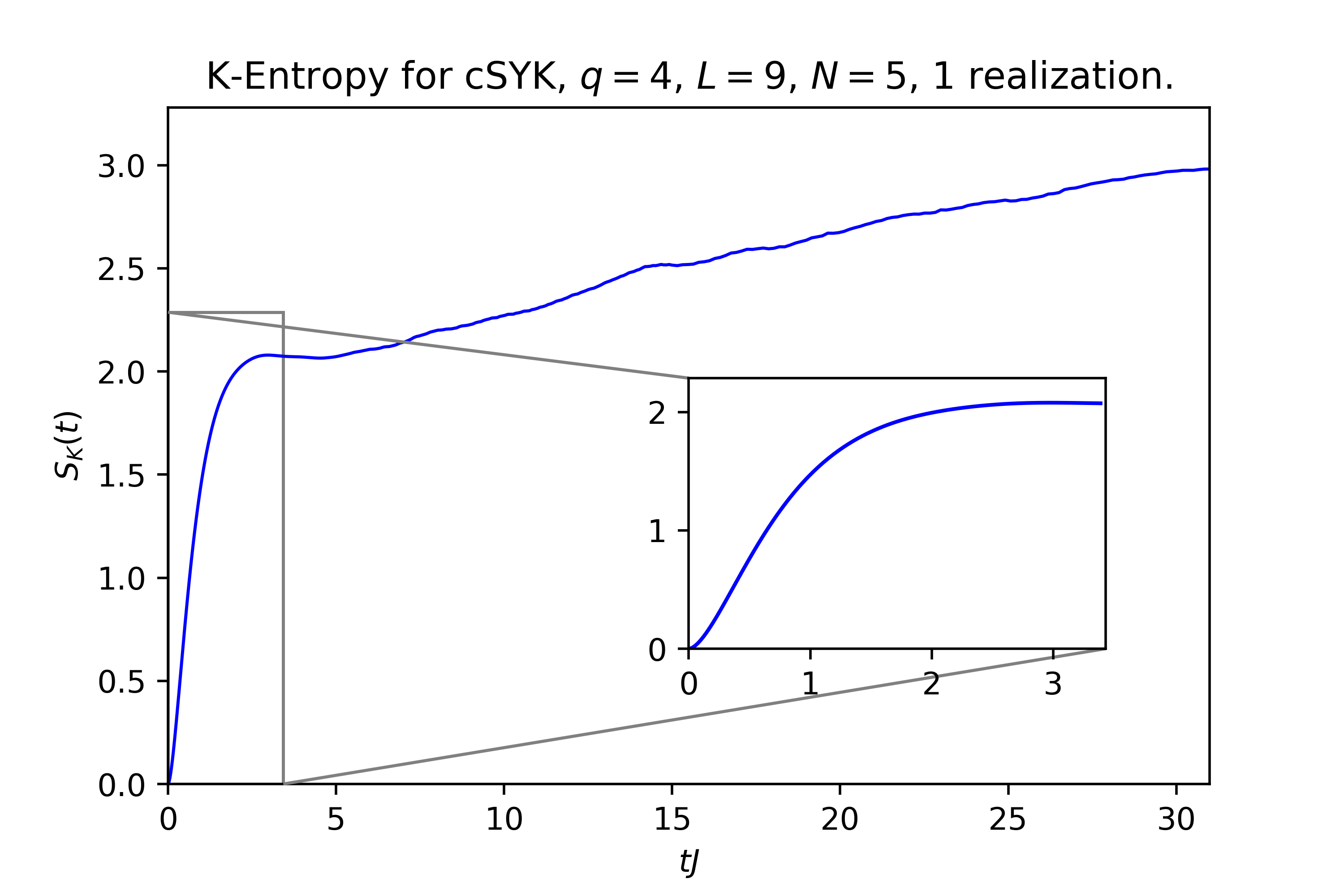}
\end{minipage}
\caption{Results for K-entropy for $L=8,\, N=4$ and $L=9,\, N=5$. \textbf{Top row:} Results for $L=8$; for exponentially long time scales in linear (left panel) and logarithmic (middle panel) scale along the horizontal axis, and for early times (right panel). The saturation value is near $L=8$.  \textbf{Bottom Row:} Results for $L=9$; for exponentially long time scales in linear (left panel) and logarithmic (middle panel) scale along the horizontal axis, and for early times (right panel). The saturation value is near $L=9$.}
\label{KS_8_9_Comp}
\end{figure*}

\begin{figure*}
    \centering 
    \begin{minipage}{.4\textwidth}
    \includegraphics[width=1.\linewidth]{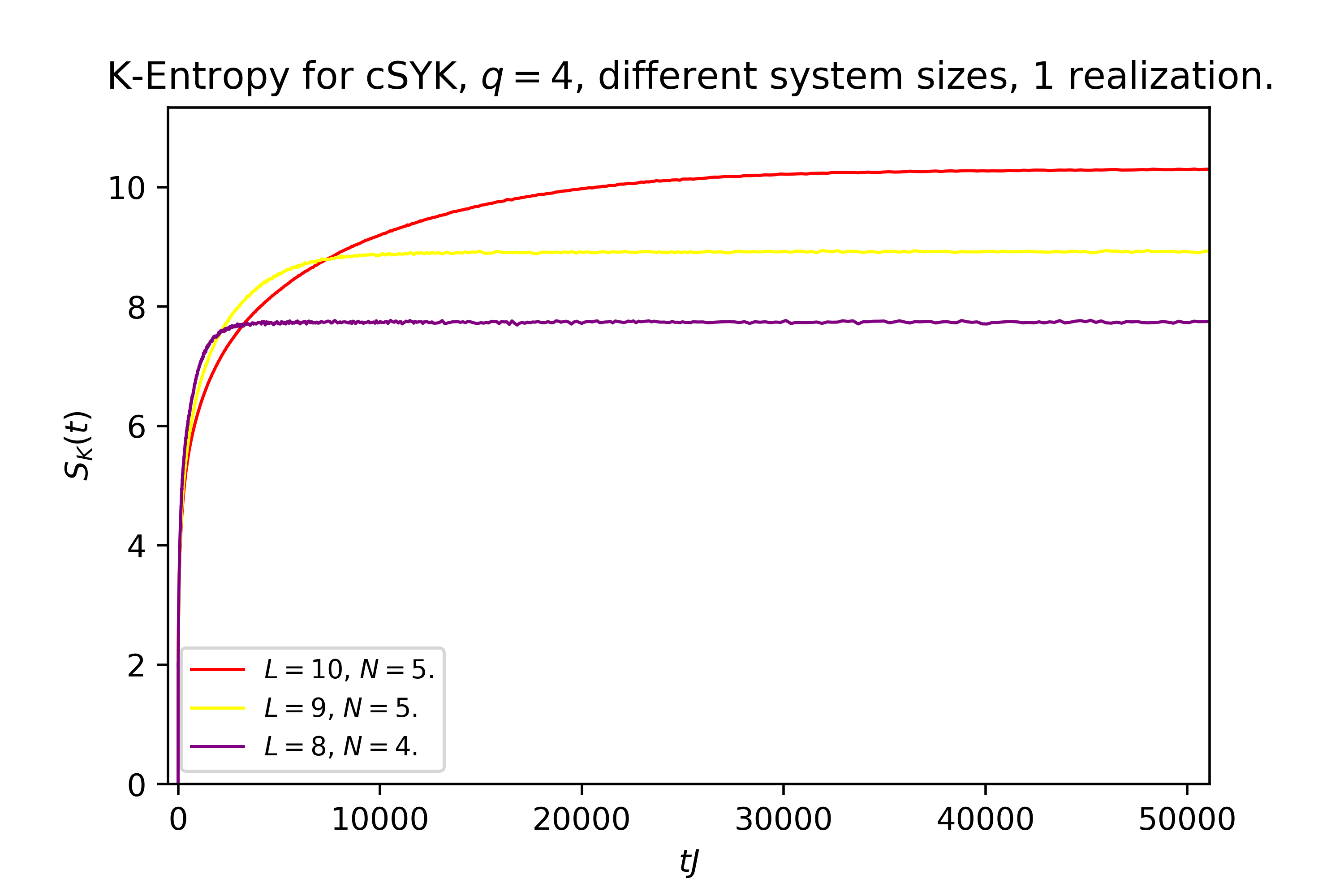}
    \end{minipage} \quad
    \begin{minipage}{.4\textwidth}
    \includegraphics[width=1.\linewidth]{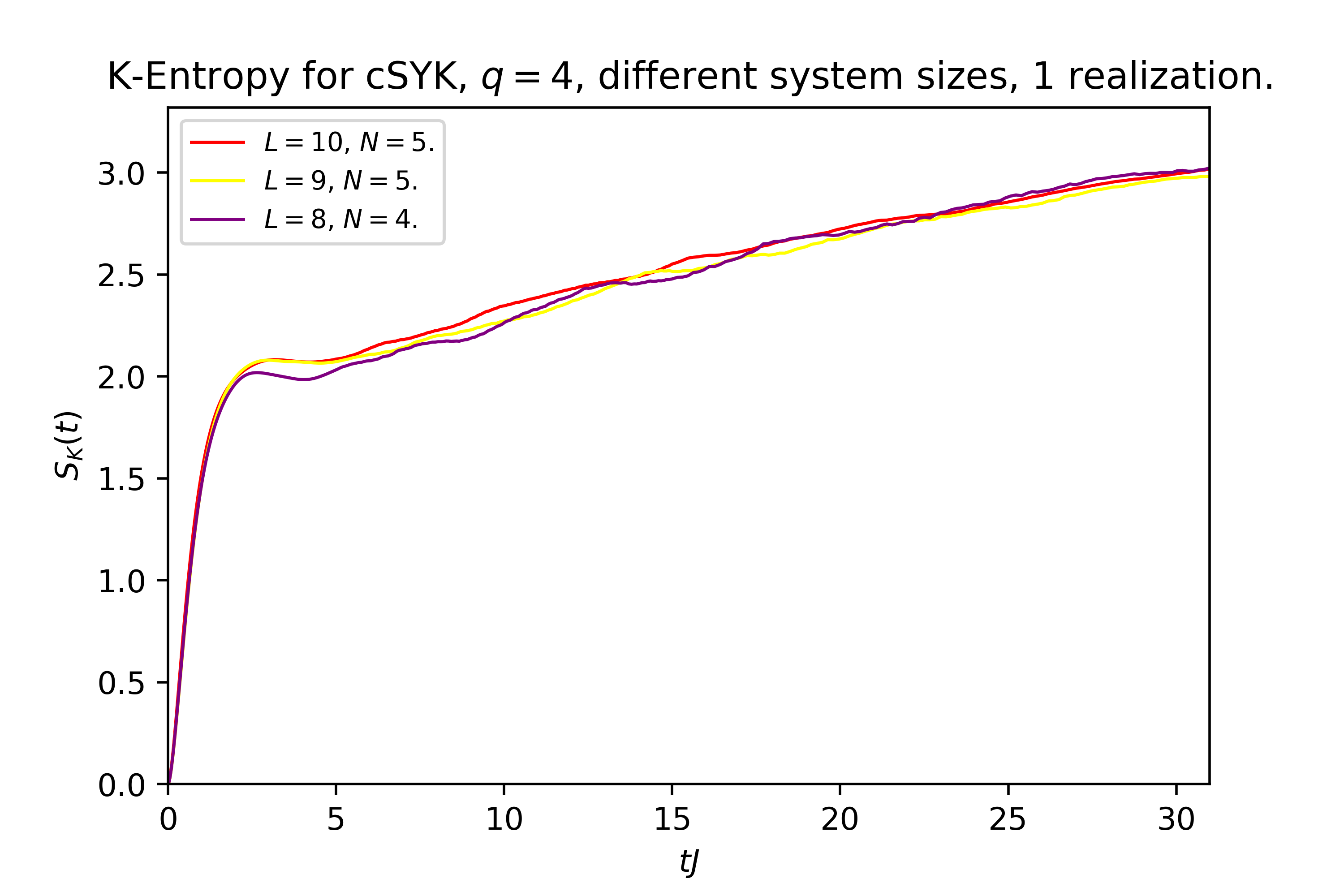}
    \end{minipage}
    \caption{Comparison of results for K-entropy for $L=8,9,10$; for long time scales (left panel), and for early times (right panel).}
    \label{KS_Size_Comp}
\end{figure*}

We studied samples with $L=8,9$ and $10$ sites. The numerical results for SYK$_4$ with $L=10$ are shown in Figures \ref{b-L10-main}, \ref{KC-L10-main} and \ref{KS-L10-main}. For details on the scaling properties of the $b$-sequence, K-complexity and K-entropy with system size, see Figures \ref{b-sequences_8_9_Comp}, \ref{KC_8_9_Comp}, \ref{KS_8_9_Comp} and \ref{KS_Size_Comp}, as well as the summarizing Table \ref{tab:summary}.


The global picture emerging from these results can be summarized in the following bullet points:

\begin{itemize}
    \item \textbf{The Descent and its associated Lanczos sequence.} The length of the Lanczos sequence saturates its upper bound. It features a period of initial growth up to $n\sim S$, followed by a regime of slow decrease to zero with roughly constant negative non-perturbative slope of order $\sim - e^{-2S}$, the Descent.
    
    \item \textbf{K-complexity} features a transition from exponential growth at very early times to linear increase. At exponentially late times it saturates at half of the Krylov dimension, since by then the operator is uniformly distributed over the Krylov basis.
    \item \textbf{K-entropy} grows linearly up to scrambling time, and then transitions to a logarithmic growth phase that continues until saturation around $S_K\sim S$ at exponentially late times.
\end{itemize}
	
The relationship between the Lanczos sequence $b_n$ and quantities like $C_K(t)$ and $S_K(t)$ is highly non-linear, which is why disorder averages need to be performed with caution. The Lanczos sequence is not really a physical observable and averaging is just used as a tool to gain knowledge about the envelope of its profile. However, in order to obtain averaged K-complexity or K-entropy as a function of time, one should compute these quantities separately for each random realization of the Hamiltonian and average over the outcomes in a final step. 
Averaging over the Lanczos sequence before computing $C_K$ and $S_K$ results in a smoothing of the $b$-sequence that stops the wave packet from randomizing efficiently before reaching the edge of the Krylov chain, being therefore reflected due to the boundary conditions. These rebounds would be reflected by large, un-physical oscillations in the profiles of $C_K$ and $S_K$, an effect that has been confirmed by numerical simulations. 



\begin{table}[H]
    \centering
    \begin{tabular}{|c|c|c|c|c|c|c|}
        \hline
         $\log(L)$ & $L$ &  $D $ & $K/2 $ & $b_n$ slope & $C_K$ sat. & $S_K$ sat.  \\
         \hline
         2.07944 & 8 & 70 & 2415.5 & $-0.00026$ & 2215  & 7.7  \\
         2.19722 & 9 & 126 & 7875.5 & $- 8.6 \times 10^{-5}$ & 7254 & 8.9 \\
         2.30258 & 10 & 252 & 31626.5 & $-2.21 \times 10^{-5}$ & 29,618 & 10.3 \\
         \hline
    \end{tabular}
    \caption{Summary Table of the numerical phenomenology observed. Values averaged over several random realizations. In terms of $S$, $\log(L)\sim\log(S), L\sim S, D \sim e^S$ and $K\sim e^{2S}$.}
    \label{tab:summary}
\end{table}

\section{Discussion} \label{Sec: Discussion}
On the analytical side, we have found an algorithmic expression for the dimension of Krylov space $K$. This quantity is very sensitive to the degeneracy structure of the spectrum and is always strictly smaller than the dimension of operator space $D^2$, since it satisfies $K\leq D^2-D+1$. Typical operators in chaotic systems are expected to saturate this bound, due to the absence of degeneracies, whereas in integrable systems $K$ will typically be significantly smaller than its upper bound (at least effectively up to late times), as verified analytically for SYK$_2$. Furthermore, numerical observations in SYK$_4$ and RMT suggest that the $b_n$-profile featuring a slow decrease to zero given by a non-perturbative slope of order $e^{-2S}$ -- the Descent -- may be a generic feature of quantum chaotic systems. It is worth noting that, since K-complexity is nothing but an average position in Krylov space, the above bound on $K$ translates directly into an upper bound for $C_K(t)$. We note that although K-complexity features the profile expected from circuit complexity, there are notable differences between them: Krylov complexity does not depend on an arbitrary tolerance parameter, and it does depend on both the Hamiltonian of the system and the operator whose K-complexity is measured over time. This may have impact in situations that involve more than one operator insertion. We hope to address this and other related issues in the future.

We end by returning to an issue we emphasized at the outset, namely that K-complexity is naturally bounded. This is a direct consequence of the finite dimensionality of the Hilbert space, and thus has the same origin as the plateau in the spectral form factor and correlation functions. We conjecture that other late-time universal characteristics of quantum chaos, namely the dip-ramp-plateau structure \cite{Cotler:2016fpe}, also leave their imprint on K-complexity. The putative bulk realization of K-complexity should therefore be sensitive to bulk Euclidean wormholes and baby universes \cite{Saad:2019lba}, in ensemble-averaged systems, such as SYK, as well as in individual quantum chaotic systems, as has been emphasized in \cite{Altland:2020ccq}.\\

{\it Note added: in the final stages of preparation of this paper, the work \cite{Jian:2020qpp} appeared on the arxiv studying, among other things, some aspects of K-complexity in SYK. Up to the timescales they consider, our results are consistent with theirs.}

\acknowledgments
We would like to thank Jos\'{e} Barb\'{o}n as well as D. Aharonov, M. Ben-Or, N. Katz, P. Nayak, H. Neuberger, J. Rougemont and M. Vielma for enlightening discussion. ER would like to thank NHETC at Rutgers Physics Department and CCPP at NYU for hospitality. The numerical computations in this paper were performed on the Baobab HPC cluster at the University of Geneva and on the Landau cluster at the Hebrew University. This work has been supported by the SNF through Project Grants 200020\_ 182513, the NCCR 51NF40-141869 The Mathematics of Physics (SwissMAP). The work of ER and RS is partially supported by the Israeli Science Foundation Center of Excellence. We thank the organisers of the conference ``Frontiers in Holography" (May 2020, Moscow), where the results of this work were presented.

\appendix

\section{Lanczos sequence in RMT}\label{Appx-RMT}
\begin{figure*}
    \begin{minipage}{.44\textwidth}
    \includegraphics[width=1.\linewidth]{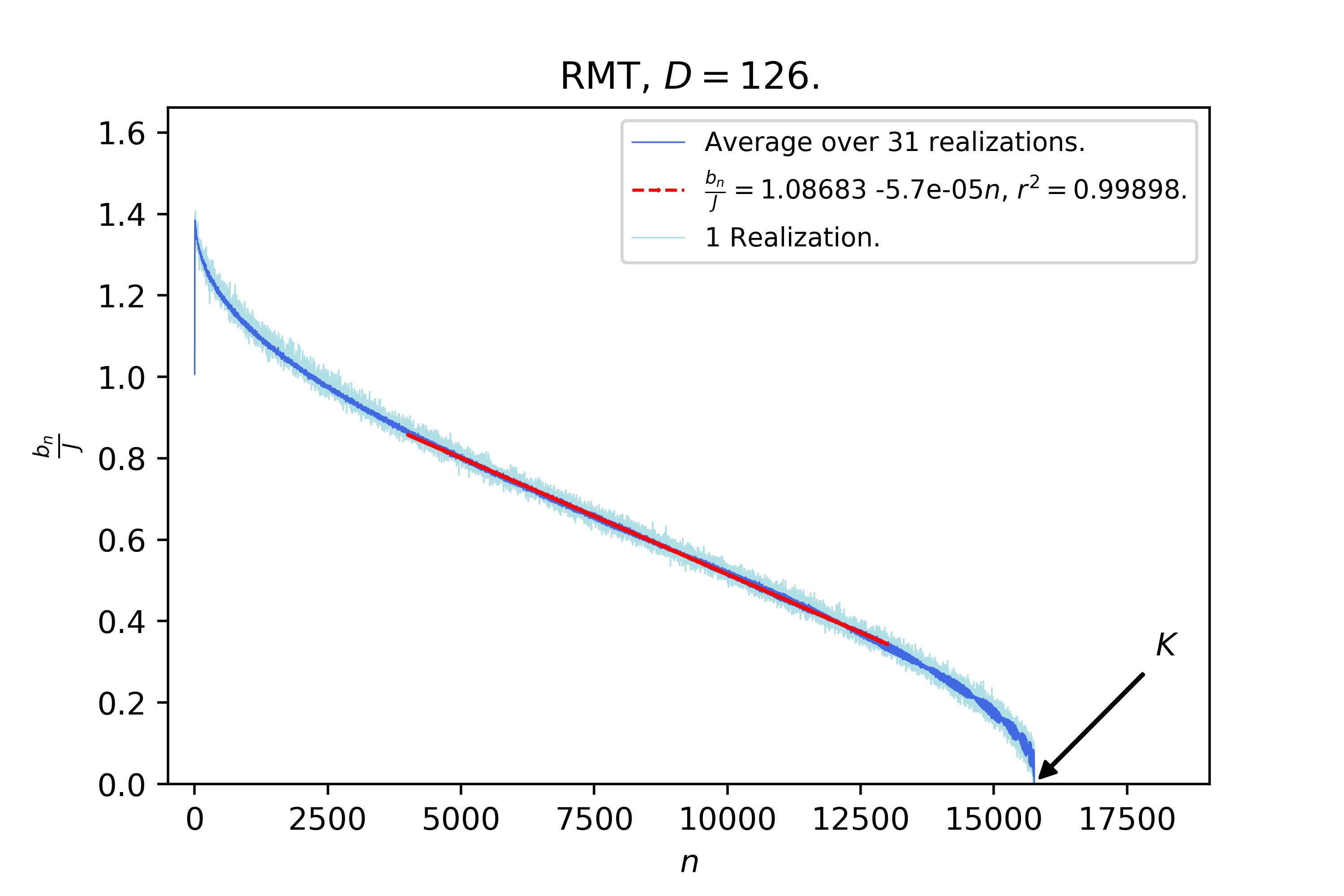}
    \end{minipage} \quad
    \begin{minipage}{.44\textwidth}
    \includegraphics[width=1.\linewidth]{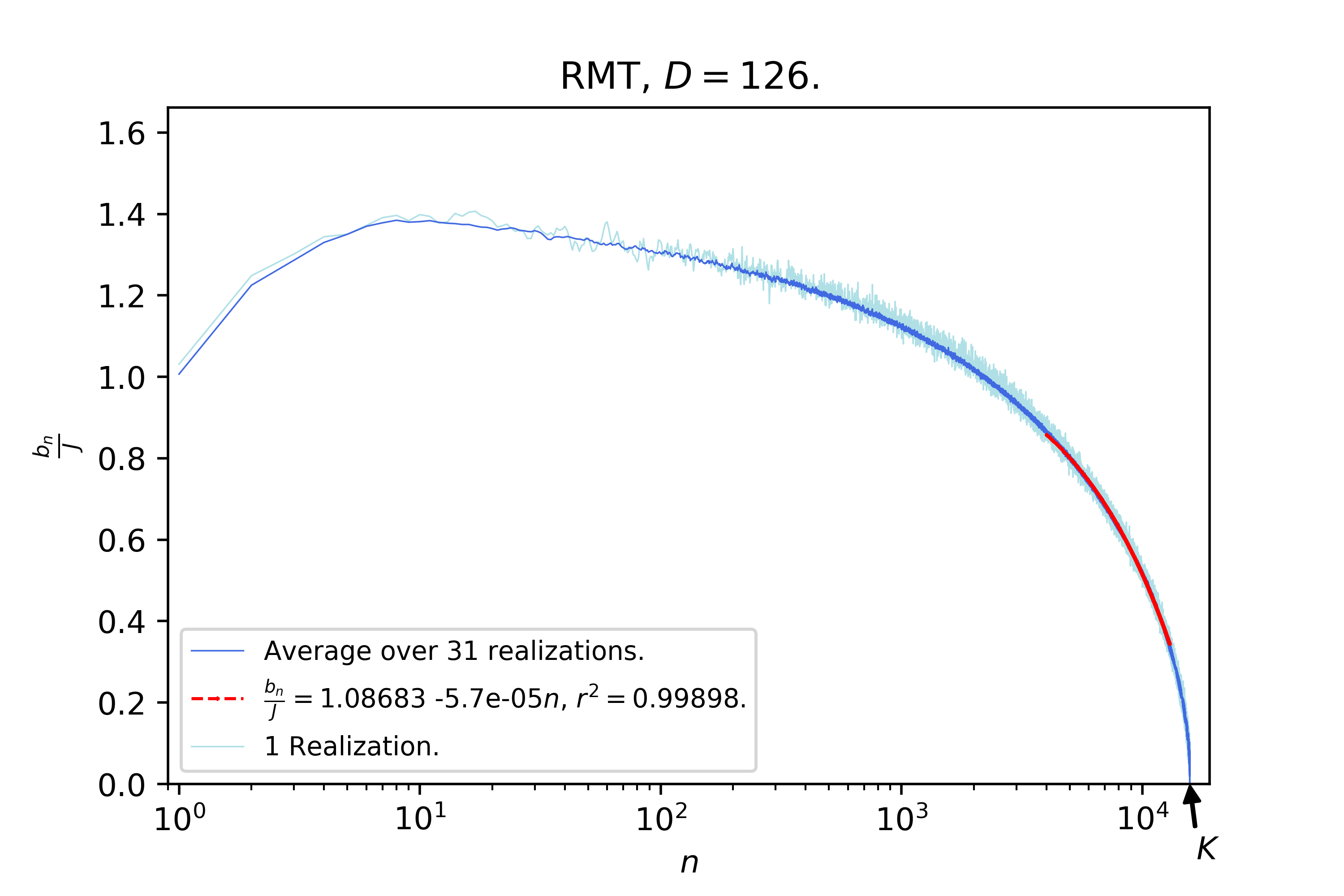}
    \end{minipage}
    \caption{Lanczos sequence for RMT drawn from the GUE with potential (\ref{RMT-GUE}). The Hilbert space dimension was chosen to match that of cSYK with $L=9$ and $N=5$. The right panel features the same data, with logarithmic scale along horizontal axis, allowing more resolution on the initial part of the sequence.}
    \label{b-RMT-D126}
\end{figure*}

Some preliminary numerical checks have indicated that RMT reproduces qualitatively the features observed in complex SYK with $q=4$, that is: saturation of the upper bound for Krylov space, $K=D^2-D+1$, and slow decrease of the $b$-sequence to zero after initial growth, with a non-perturbative slope of order $\sim-\frac{1}{K}\sim-e^{-2S}$, $S$ being the entropy (system size). Figure \ref{b-RMT-D126} depicts the Lanczos sequence of a system whose Hamiltonian $H$ is drawn from a GUE with potential:
\begin{equation}
\centering
\label{RMT-GUE}
V(H) = \frac{D}{2J^2}\text{Tr}\left(H^2\right)
\end{equation}
where $D$ is the Hilbert space dimension and $J$ is the coupling strength that sets energy scales. A more detailed numerical and analytical study of the Lanczos sequence in RMT constitutes work in progress, but these preliminary checks make it natural to conjecture that the discussed features are universal in chaotic systems.

\section{Krylov space for SYK$_2$}\label{Appx-SYK2}
Given the SYK$_2$ model described in \ref{MajoranaSYK-q2}, we can proceed to apply the Lanczos algorithm analytically to the operator $\mathcal{O}=\chi_A$.

We will use the Frobenius scalar product defined in the main text, recalling that the dimension of the Hilbert space of states in this case is $D=2^M$. It is not difficult to prove that:
\begin{equation}
    \centering
    \label{Majoranas-overlap}
    \left(\chi_i|\chi_j\right) = \frac{1}{2}\delta_{ij} ~.
\end{equation}

\begin{itemize}
    \item $\mathcal{O}_0$:
    Making use of (\ref{Majoranas-overlap}) one finds that our starting operator is not normalized, since $\left(\mathcal{O}|\mathcal{O}\right)=\frac{1}{2}$, so:
    \begin{equation}
        \centering
        \label{q2-O0}
        \mathcal{O}_0 = \frac{\mathcal{O}}{\sqrt{\left(\mathcal{O}|\mathcal{O}\right)}} = \sqrt{2}\mathcal{O}=\sqrt{2}\,\chi_A ~.
    \end{equation}
    
    \item $\mathcal{O}_1$:
    One first needs to compute $\mathcal{A}_1$:
    \begin{equation}
        \centering
        \label{q2-A1}
        \mathcal{A}_1 = \left[H,\mathcal{O}_0\right] = i\frac{\sqrt{2}}{2}\sum_{i,j=1}^{L=2M}m_{ij}\left[\chi_i\chi_j,\chi_A\right]
    \end{equation}
    We make use of the commutator of the fermionic bilinear with the single Majorana:
    \begin{equation}
        \centering
        \label{q2-Commut}
        \left[\chi_i\chi_j,\chi_A\right] = \delta_{Aj}\chi_i-\delta_{Ai}\chi_j
    \end{equation} 
    so that finally
    \begin{equation}
        \centering
        \label{q2-A1-final}
        \mathcal{A}_1=\left[H,\mathcal{O}_0\right] = i\sqrt{2}\sum_{\substack{i = 1 \\ i\neq A}}^{L=2M}m_{iA}\chi_i ~.
    \end{equation}
    
    The first Lanczos coefficient is:
    \begin{equation}
        \centering
        \label{q2-b1}
        b_1 = \sqrt{\left(\mathcal{A}_1|\mathcal{A}_1\right)}=\sqrt{2\sum_{\substack{i,j=1 \\ i\neq A \\ j\neq A}}^{L=2M}m_{iA}m_{jA}\left(\chi_i|\chi_j\right)}
    \end{equation}
    and recalling (\ref{Majoranas-overlap}) one finds:
    \begin{equation}
        \centering
        \label{q2-b1-final}
        b_1 = \sqrt{\sum_{\substack{i=1 \\ i\neq A}}^{L=2M}m_{iA}^2} ~.
    \end{equation}
    Thus, the next Krylov element is:
    \begin{equation}
        \centering
        \label{q2-O1}
        \mathcal{O}_1 = \frac{1}{b_1}\mathcal{A}_1 = \frac{i\sqrt{2}}{\sqrt{\mathlarger{\sum}_{\substack{i=1 \\ i\neq A}}^{L=2M}m_{iA}^2}} \sum_{\substack{i=1 \\ i\neq A}}^{L=2M}m_{iA}\,\chi_i ~.
    \end{equation}
    The form of (\ref{q2-b1-final}) allows the computation of the ensemble average of the first Lanczos coefficient, since the calculation will turn out to be simple if one makes use of spherical coordinates in sample space:
    \begin{equation}
        \centering
        \label{q2-Eb1}
        \mathbb{E}\left(b_1\right) = \int d^{L-1} x\, b(\mathbf{x}) P(\mathbf{x})
    \end{equation}
    where we defined $b(\mathbf{x})\equiv\sqrt{\mathbf{x}^2}$, and note that there are $L-1$ random variables because (\ref{q2-b1-final}) only involves $L-1$ independent coupling strengths. The probability distribution is a product of Gaussians:
    \begin{equation}
        \centering
        \label{q2-b1-pdf}
        P(\mathbf{x}) = \prod _{i=1}^{L-1}\rho(x_i),\;\;\;\;\rho(x_i)=\frac{1}{\sigma \sqrt{2\pi}}e^{-\frac{x_i^2}{2\sigma^2}}
    \end{equation}
    where, in agreement with $ \mathbb{E} (m_{ij}^2) = m^2/L$, the standard deviation is given by $\sigma = \frac{m}{\sqrt{L}}$.
    As promised, the use of spherical coordinates greatly simplifies the computation and the result is:
    \begin{equation}
        \centering
        \label{q2-Eb1-final}
        \mathbb{E}\left(b_1\right) = \frac{\Gamma\left(\frac{L}{2}\right)}{\Gamma\left(\frac{L-1}{2}\right)}\sqrt{\frac{2}{L}}m\overset{L\to +\infty}{\longrightarrow} m ~.
    \end{equation}
    
    \item \textbf{More elements:} It can be an interesting task to compute in closed form the rest of the Lanczos sequence, but we can already anticipate what its length will be: As shown in (\ref{q2-Commut}), the commutator of a bilinear with a single-site operator will still yield a linear combination of Majoranas on a single site, so $\mathcal{A}_2=\left[H,\mathcal{O}_1\right]-b_1 \mathcal{O}_0$ will still be a one-site operator, and so will the rest of the Krylov elements. Hence the maximum possible number of linearly independent Krylov elements will be given by the number of sites of the system, that is to say, in this integrable system the Krylov dimension $K$ is bounded by:
    \begin{equation}
        \centering
        \label{q2-KDIM}
        K\leq L \sim \log D \ll D^2-D+1
    \end{equation}
    i.e. the Krylov space scales at most linearly with entropy (system size), instead of with the operator space dimension (exponential in entropy).
\end{itemize}

\section{Numerical algorithms}\label{Appx-algorithms}

It is known that the original Lanczos algorithm, as presented in the main text, features an important numerical instability \cite{PRO,SO,Parlett_book}: the construction of each Krylov element makes use of the two previous ones, so errors due to finite-precision arithmetic accumulate dramatically and orthogonality of the Krylov basis is soon lost in numerical computations. Residual overlaps between Krylov elements grow exponentially (or even faster) with the iteration number $n$, which makes the Lanczos coefficients unreliable after a few iterations. In particular, the original Lanczos algorithm doesn't feature the termination of the sequence by hitting a zero at $n=K$ when run with finite precision, and instead it generally outputs a Lanczos sequence that oscillates wildly around some constant value, whose disorder average yields a completely flat sequence (after initial growth) in complex SYK; this is purely a product of numerical precision errors and needs to be corrected in order to shed light on the structure of the $b$-sequence along the full length of the Krylov chain. Some numerical algorithms need to be implemented in order to cure this instability, allowing observation of the slow decay to zero of the Lanczos coefficients after the initial growth in complex SYK. Such algorithms are described below.

\subsection{Full Orthogonalization (FO)}

This algorithm \cite{Parlett_book} performs a brute-force re-orthogonalization of the newly constructed Krylov element with respect to the previous ones at every iteration of the Lanczos algorithm, ensuring orthonormality of the Krylov basis up to machine precision $\varepsilon_M$. The algorithm is not very efficient time-wise, and is also costly in terms of memory, since the whole Krylov basis needs to be stored and is used in every iteration, it is therefore used mainly when one wants to compute only part of the Lanczos-sequence (see for example the recent work \cite{Yates_2020}). However, for small samples, FO can be used safely and one can check that the results yielded agree with the theoretical predictions described previously in this article (length of the Lanczos sequence, termination by hitting a zero, eigenvalues of the tri-diagonal matrix matching those corresponding to the eigenspace representatives that span the Krylov space). 

The FO algorithm amounts to performing explicit Gram-Schmidt at every iteration in the Lanczos algorithm to ensure orthogonality (up to machine precision). For numerical purposes, it is usually optimal to perform Gram-Schmidt \textit{twice} every time:

\begin{enumerate}
    
    \item $\left|\mathcal{O}_0\right) = \frac{1}{\sqrt{\left(\mathcal{O}|\mathcal{O}\right)}}\left|\mathcal{O}\right)$.
    \item For $n\geq 1$: Compute $\left|\mathcal{A}_n\right) = \mathcal{L}\left|\mathcal{O}_{n-1}\right)$.
    \item Re-orthogonalize $\left|\mathcal{A}_n\right)$ explicitly with respect to all previous Krylov elements: \\
    $\left|\mathcal{A}_n\right)\longmapsto \left|\mathcal{A}_n\right)-\sum_{m=0}^{n-1}\left|\mathcal{O}_m\right)\left(\mathcal{O}_m|\mathcal{A}_n\right)$.
    \item Repeat step 3.
    \item Set $b_n = \sqrt{\left(\mathcal{A}_n|\mathcal{A}_n\right)}$.
    \item If $b_n=0$ stop; otherwise set $\left|\mathcal{O}_n\right) = \frac{1}{b_n}\left|\mathcal{A}_n\right)$ and go to step 2.
\end{enumerate}

\subsection{Partial Re-Orthogonalization (PRO)}

This algorithm \cite{PRO} allows the residual overlaps between Krylov elements to grow up to a certain threshold, and re-orthogonalization is only performed when the threshold is crossed. For a machine precision $\varepsilon_M$, the threshold is typically taken to be $\sqrt{\varepsilon_M}$.

In our notation, the recursion relation for the Krylov basis is, including finite-precision errors:
\begin{equation}
\centering
\label{Lanczos-recursion}
b_{n}\left| \mathcal{O}_n \right) = \mathcal{L}\left| \mathcal{O}_{n-1} \right)-b_{n-1}\left| \mathcal{O}_{n-2} \right)+\left| \xi_{n-1} \right)
\end{equation}
where $\left| \xi_{n-1} \right)$ accounts for some spurious vector generated by accumulated numerical errors. All the objects denoted above represent the \textit{quantities numerically computed}, rather than the actual (analytically exact) Lanczos coefficients and Krylov elements. Acting with $\left( \mathcal{O}_{k} \right|$ from the left:
\begin{equation}
\centering
\label{Act-from-left}
b_{n}\left( \mathcal{O}_{k} \right.\left| \mathcal{O}_n \right) = \left( \mathcal{O}_{k} \right|\mathcal{L}\left| \mathcal{O}_{n-1} \right)-b_{n-1}\left( \mathcal{O}_{k} \right.\left| \mathcal{O}_{n-2} \right)+\left( \mathcal{O}_{k} \right.\left| \xi_{n-1} \right)
\end{equation}
where we recall that $\left( \mathcal{O}_{k} \right|\mathcal{L}\left| \mathcal{O}_{n-1} \right) = T_{k,n-1}$, $T$ being the tri-diagonal symmetric matrix built out of the Lanczos sequence, see (\ref{L-Tridiagonal}). We now define the matrix $W$, with items $W_{kn}=\left( \mathcal{O}_{k} \right.\left| \mathcal{O}_n \right)$, that is:
\begin{equation}
\centering
\footnotesize
\label{w-items}
\left(W_{kn}\right)=\begin{pmatrix}
W_{00}=\left( \mathcal{O}_{0} \right.\left| \mathcal{O}_0 \right)&W_{01}=\left( \mathcal{O}_{0} \right.\left| \mathcal{O}_1 \right)&W_{02}=\left( \mathcal{O}_{0} \right.\left| \mathcal{O}_2 \right)&\dots\\
\,&W_{11}=\left( \mathcal{O}_{1} \right.\left| \mathcal{O}_1 \right)&W_{12}=\left( \mathcal{O}_{1} \right.\left| \mathcal{O}_2 \right)&\dots\\
\,&\,&W_{22}=\left( \mathcal{O}_{2} \right.\left| \mathcal{O}_2 \right)&\dots\\
\vdots&\vdots&\vdots&\ddots
\end{pmatrix}
\end{equation}

Even though we don't write all the entries in the matrix, it's hermitian by definition of the scalar product $\left(\cdot | \cdot \right)$, so $W=W^\dagger\Longleftrightarrow W_{kn}=W_{nk}^{*}$. In the PRO algorithm, however, we construct iteratively the entries written explicitly in (\ref{w-items}): For a given $n$ we'll want to estimate $W_{kn}$, for $k\leq n$. We do so by noticing that (\ref{Act-from-left}) is nothing but:
\begin{equation}
\centering
\label{Act-from-left-w}
b_{n} W_{kn} = T_{k,n-1} -b_{n-1} W_{k,n-2} + \left( \mathcal{O}_{k} \right.\left| \xi_{n-1} \right).
\end{equation}
Renaming indices $k\leftrightarrow n-1$ (i.e. $k\mapsto n-1$ and $n\mapsto k+1$):
\begin{equation}
\centering
\label{Act-from-left-w-rename-ind}
b_{k+1} W_{n-1,k+1} = T_{n-1,k} -b_{k} W_{n-1,k-1} + \left( \mathcal{O}_{n-1} \right.\left| \xi_{k} \right).
\end{equation}

Computing $(\ref{Act-from-left-w})-(\ref{Act-from-left-w-rename-ind})$, recalling that $T$ is symmetric and solving for $W_{kn}$:
\begin{equation}
\centering
\label{Recursion-w}
\begin{split}
W_{kn} & =\frac{1}{b_n}\left[b_{k+1} W_{k+1,n-1}^{*} + b_k W_{k-1,n-1}^{*}-b_{n-1} W_{k,n-2}\right. \\ 
& +\left. \left( \mathcal{O}_{k} \right.\left| \xi_{n-1} \right)- \left( \mathcal{O}_{n-1} \right.\left| \xi_{k} \right)\right] .
\end{split}
\end{equation}

We want to use (\ref{Recursion-w}) to determine, for a fixed $n$, all $W_{kn}$ with $k\leq n$ given that we know all $\left\{ W_{ij},\;i=0,...,j,\;\forall j=0,...,n-1 \right\}$ (according to what we depicted in (\ref{w-items}): we compute iteratively each upper-diagonal column making use of the previous ones). 

We note that $W_{nn}$ is not determined by (\ref{Recursion-w}) in terms of previous upper-diagonal columns, but we can set it to $W_{nn}=1$ because in the $n$-th Lanczos step, $\left| \mathcal{O}_n \right)$ is explicitly normalized to unity. Likewise, $W_{n-1,n}$ is not determined from (\ref{Recursion-w}) in terms of previous columns, and the Lanczos recursion (\ref{Lanczos-recursion}) doesn't explicitly orthogonalize $\left| \mathcal{O}_n \right)$ against $\left| \mathcal{O}_{n-1} \right)$, so we will need to orthogonalize them explicitly, and then set $W_{n-1,n}=\mathit{O}\left(\varepsilon_M\right)$ (i.e. zero up to machine precision).

Also, an estimate for $\left( \mathcal{O}_{k} \right.\left| \xi_{n-1} \right)- \left( \mathcal{O}_{n-1} \right.\left| \xi_{k} \right)$ is needed. One can take something of order of the machine precision times the norm of the Liouvillian:
\begin{equation}
\centering
\label{estimate}
\Big{|}\left( \mathcal{O}_{k} \right.\left| \xi_{n-1} \right)- \left( \mathcal{O}_{n-1} \right.\left| \xi_{k} \right)\Big{|} \sim 2\varepsilon_M\|\mathcal{L}\|
\end{equation}
where $\big{|}\big{|}\mathcal{L}\big{|}\big{|}$ should be the norm of the Liouvillian induced by the scalar product $\left(\cdot | \cdot\right)$ in the Hilbert space of operators. For practical applications, this contribution can just be ignored, since in any case each iteration of the algorithm will already generate spurious errors of order $\varepsilon_M$.

All in all, the LanPRO algorithm reads:
\begin{itemize}
	\item Compute $\left|\mathcal{O}_{0}\right)=\frac{1}{\sqrt{\left(\mathcal{O}\right.\left|\mathcal{O}\right)}}\left|\mathcal{O}\right)$.
	\begin{itemize}
		\item Set $W_{00}=1$.
	\end{itemize}
	\item Compute $\left|\mathcal{A}_{1}\right) = \mathcal{L}\left|\mathcal{O}_{0}\right)$. 
		
		\begin{itemize}
		
		\item Orthogonalize it explicitly with respect to $\left|\mathcal{O}_{0}\right)$.
		\item Compute $	b_1=\sqrt{\left(\mathcal{A}_1|\mathcal{A}_1\right)}$. \textbf{If} $b_1<\sqrt{\varepsilon_M}$ \textbf{stop.} Otherwise compute $\left| \mathcal{O}_1 \right)=\frac{1}{b_1}\left|\mathcal{A}_{1}\right) $.
		\item Set $W_{01}=\varepsilon_M$ and $W_{11}=1$.
		\end{itemize}
	
	\item \textbf{Loop} for $n\geq 2$, and for every $n$ \textbf{do}:
	\begin{itemize}
		\item Compute $\left|\mathcal{A}_{n}\right) = \mathcal{L}\left|\mathcal{O}_{n-1}\right)-b_{n-1}\left|\mathcal{O}_{n-2}\right)$.
		\item Compute the \textit{a-priori Lanczos coefficient:} \\ $b_n=\sqrt{\left(\mathcal{A}_n|\mathcal{A}_n\right)}$.
		\item \textbf{If} $b_n<\sqrt{\varepsilon_M}$ \textbf{break}, otherwise continue...

		\item Orthogonalize explicitly $\left| \mathcal{A}_n \right)$ with respect to $\left| \mathcal{O}_{n-1}\right)$.
		\item Set $W_{n,n}=1$ and $W_{n-1,n}=\varepsilon_M$.
		\item \textbf{Loop} for all $k=0,...,n-2$, determine $W_{kn}$ \textbf{doing}:
		\begin{equation}
		\centering
		\label{Assign-w}
		\begin{split}
		&\widetilde{W} = b_{k+1} W_{k+1,n-1}^{*} + b_k W_{k-1,n-1}^{*}-b_{n-1} W_{k,n-2}  \\
		&W_{kn}=\frac{1}{b_n}\left[\widetilde{W}+\frac{\widetilde{W}}{\big|\widetilde{W}\big|}\cdot 2\varepsilon_M\|\mathcal{L}\| \right]
		\end{split}	
		\end{equation}
		
		\item\textbf{If} there is some $k\leq n-2$ such that $W_{kn}>\sqrt{\varepsilon_M}$,  \textbf{do}: 
		\begin{itemize}
			\item Re-orthogonalize explicitly $\left| \mathcal{A}_n \right)$ and $\left| \mathcal{A}_{n-1} \right)$ with respect to all previous Krylov elements.
			\item From the new $\left|\mathcal{A}_{n-1}\right)$, re-compute $b_{n-1}$. \textbf{Break} if $b_{n-1}<\sqrt{\varepsilon_M}$, \textbf{otherwise} re-compute $\left|\mathcal{O}_{n-1}\right)$.
			\item From the new $\left|\mathcal{A}_{n}\right)$, re-compute $b_{n}$. \textbf{Break} if $b_{n}<\sqrt{\varepsilon_M}$, \textbf{otherwise} compute $\left|\mathcal{O}_{n}\right)=\frac{1}{b_n}\left|\mathcal{A}_n\right)$.
			\item Set $W_{a,n-1}=\delta_{a,n-1}$+$\left(1-\delta_{a,n-1}\right)\varepsilon_M$, for all $a=0,...,n-1$.
			\item Set $W_{a,n}=\delta_{a,n}$+$\left(1-\delta_{a,n}\right)\varepsilon_M$, for all $a=0,...,n$.
		\end{itemize}
		\item \textbf{Otherwise} compute  $\left|\mathcal{O}_{n}\right)=\frac{1}{b_n}\left|\mathcal{A}_n\right)$.
		
	\end{itemize}
 
\end{itemize}

(End of algorithm).\\

Some comments are in order:

\begin{itemize}
	\item In every iteration $n\geq 2$, only the two previous upper-diagonal columns $\left\{W_{k,n-1},\;k=0,...,n-1\right\}$ and $\left\{W_{k,n-2},\;k=0,...,n-2\right\}$ of the matrix $W$ are required. This is why, whenever re-orthogonalization is required, one only carries it out for $\left|\mathcal{A}_{n-1}\right)$ and $\left|\mathcal{A}_{n}\right)$.
	\item Whenever re-orthogonalization is needed, it is optimal to perform it twice, in the same way Gram-Schmidt is applied twice at every step of the FO algorithm.
\end{itemize}

In the case of complex SYK$_4$, for the system sizes studied in this article and a typical floating point precision of $\varepsilon_M\sim 10^{-15}$, PRO reduces the number of re-orthogonalizations required by approximately a factor of $10$, as compared to FO.

\bibliographystyle{unsrt}
\bibliography{main}

\end{document}